\documentclass[sigconf]{acmart}
\usepackage{placeins,acmart-taps}
\usepackage{tabularx}
\usepackage{alltt}

\AtBeginDocument{%
  }

\copyrightyear{2026}
\acmYear{2026}
\setcopyright{cc}
\setcctype{by}
\acmConference[CHIWORK '26]{Proceedings of the 5th Annual Symposium on Human-Computer Interaction for Work}{June 22--25, 2026}{Linz, Austria}
\acmBooktitle{Proceedings of the 5th Annual Symposium on Human-Computer Interaction for Work (CHIWORK '26), June 22--25, 2026, Linz, Austria}
\acmDOI{10.1145/3808045.3808071}
\acmISBN{979-8-4007-2429-9/2026/06}

\usepackage{ragged2e}
\usepackage{fvextra}

\usepackage{framed}

\DefineVerbatimEnvironment{mdcode}{Verbatim}{
  samepage=false,
  breaklines=true,
  breakanywhere=true,
  breakbefore={_./-},
  breakafter={_./-,},
  breaksymbol={},
  breaksymbolleft={},
  breaksymbolindent=0pt,
  obeytabs=true, tabsize=2, showtabs=false,
  formatcom={\setlength{\parindent}{0pt}\RaggedRight},
  xleftmargin=0pt,
  framesep=0pt,
  fontsize=\small,
  listparameters={
    \setlength{\topsep}{0pt}
    \setlength{\partopsep}{0pt}
    \setlength{\parskip}{0pt}
  }
}

\begin{document}

\title{Understanding, Challenging, and Demystifying Perceptions of Gig Worker Vulnerabilities}

\author{Sander de Jong}
\orcid{0000-0002-2591-3805}
\affiliation{%
  \institution{Aalborg University}
  \city{Aalborg}
  \country{Denmark}
}
\email{sanderdj@create.aau.dk}

\author{Jane Hsieh}
\orcid{0000-0002-4933-1156}
\affiliation{%
  \institution{Carnegie Mellon University}
  \city{Pittsburgh}
  \state{PA}
  \country{USA}
}
\email{jhsieh2@andrew.cmu.edu}

\author{Tzu-Sheng Kuo}
\orcid{0000-0002-1504-7640}
\affiliation{%
  \institution{Carnegie Mellon University}
  \city{Pittsburgh}
  \state{PA}
  \country{USA}
}
\email{tzushenk@cs.cmu.edu}

\author{Rune Møberg Jacobsen}
\orcid{0000-0002-1877-1845}
\affiliation{%
  \institution{Aalborg University}
  \city{Aalborg}
  \country{Denmark}
}
\email{runemj@cs.aau.dk}

\author{Niels van Berkel}
\orcid{0000-0001-5106-7692}
\affiliation{%
  \institution{Aalborg University}
  \city{Aalborg}
  \country{Denmark}
}
\email{nielsvanberkel@cs.aau.dk}

\author{Haiyi Zhu}
\orcid{0000-0001-7271-9100}
\affiliation{%
  \institution{Carnegie Mellon University}
  \city{Pittsburgh}
  \state{PA}
  \country{USA}
}
\email{haiyiz@cs.cmu.edu}

\newcommand{\jh}[1]{\textcolor{violet}{[JH: #1]}}
\renewcommand{\shortauthors}{S. de Jong et al.}

\begin{abstract}
Across service domains, platform-based gig workers often face a wide range of severe yet hidden vulnerabilities, including opaque pay practices, illusions of flexibility, health and safety risks, and privacy violations. To the general public and inexperienced workers such latent vulnerabilities remain largely unknown and concealed by intentional platform design that gives illusions of adequate labor protections, or \textit{myths}.
This study examines how workers perceive (and shift their beliefs away from) five commonly held misconceptions regarding gig worker vulnerabilities.
In Phase~I, crowdworkers ($N~=~236$) rated their agreement with five common myths surrounding vulnerabilities in gig work:$~227$ of them believed one or more myth(s).
In Phase~II, we challenged these workers to defend their views by presenting an expert- or LLM-generated counterargument. 
Our findings show workers' underexposure to personal and shared vulnerabilities of gig work, revealing a knowledge gap where persuasive interventions can scalably raise awareness around such hidden labor conditions.
We reflect on the effectiveness of different persuasion strategies and discuss implications for promoting more accurate public perceptions that support collective bargaining of workers' rights.
\end{abstract}




\maketitle

\section{Introduction}
Legally classified as independent contractors, platform-based gig workers operate on their own accord and often lack physical spaces and digital infrastructures that mediate watercooler conversations and social support around their working conditions~\cite{Yao2021TogetherAlone, networked}.
For individual workers, such absence of knowledge exchange exacerbates information asymmetries \cite{asymmetry} -- giving rise to incomplete understandings and even myths surrounding how platforms function, which shape working conditions and vulnerabilities~\cite{vul_2}.
Information asymmetries are further influenced by how crowd work systems balance anonymity and transparency. While greater transparency can support accountability and coordination~\cite{Huang2013DontHideCrowd}, anonymity can encourage more active participation and higher-quality feedback~\cite{Hui2015UsingAnonymityCommunal}. Together, these dynamics determine whether workers share detailed accounts or limited information, constraining the visibility of working conditions.
For the worker collective, the lack of peer-to-peer information flows hampers their opportunities to mobilize and coordinate action for advancing labor rights~\cite{Wu2022ReasonableThing, dasgupta2025systematic, dalalthesis}.

A growing body of research \cite{paranoia, ticona2019beyond, beyond2, Yao2021TogetherAlone, privacy, speculative, Hsieh2023GigWorkersWellbeing} -- and, more recently, informal knowledge from online discussion forums \cite{quallm} -- document a wide range of severe but often obscured vulnerabilities in gig workers’ conditions (\textit{e.g.,} opaque pay practices, health and safety risks, privacy violations). These vulnerabilities remain largely unknown to the general public and inexperienced workers, in part because platforms intentionally design out communication structures between workers and clients \cite{asymmetry}, strategically manage discourse during public-facing crises \cite{crisis} and maintain limited data transparency \cite{Rao2024Rideshare}. Additionally, workers may develop folk theories around the workings of platforms' algorithmic infrastructures, creating everyday explanations that are shaped by limited information and feedback rather than direct knowledge of system design~\cite{Eslami2016FirstItThen, Karizat2021AlgorithmicFolkTheories}. Together, these practices foster inaccurate perceptions and misheld beliefs, which we characterize as \textbf{\textit{myths}} in this work.

Our paper focuses on five common myths in gig work. The first issue centres the low and unequal wages that historically drove workers to protest and unionize for better treatment. In the case of platform-based gig work, algorithmic wage discrimination systematically codifies and amplifies the inequality~\cite{gender_platform, dubal2023algorithmic, Airbnb}, but inexperienced workers often remain unaware of their \textit{financial precarity}. Second, many underlying \textit{health and safety hazards} are not immediately apparent to new gig workers but can emerge over time after cumulative physical strain~\cite{fatigue} or compounding psychological stress~\cite{longitudinal, hazards}. Third, \textit{privacy intrusions} and surveillance are also common but covert algorithmic management practices that threaten worker autonomy, agency and psychological well-being~\cite{privacy, surveillance, watched, watching}.  Fourth, companies heavily market schedule flexibility as an affordance of platform-based work, with Uber once bearing the slogan\footnote{https://www.uber.com/us/en/drive/}  ``\textit{Drive when you want. Earn what you need}.'' But in practice, this fabricated \textit{illusion of flexibility} is met with contrasting realities of low work availability~\cite{flexibility}, intense but manufactured competition~\cite{rock}, as well as long and irregular working hours \cite{good}. Finally, workers' classification as independent contractors~\cite{war} deprives them of standard worker protections (\textit{e.g.,} minimum wage, health insurance), although many are not consciously aware of their own \textit{exclusion from labor rights and protections}. Addressing this awareness gap requires interventions beyond passive provisions of information. Instead, workers should actively engage and reflect on their own beliefs in a scalable, yet personalised  manner.

Recent LLM advances offer a potential pathway for building such interventions. State-of-the-art models can formulate arguments in a very human-like and persuasive manner~\cite{Bender2021StochasticParrots, Jones2024LiesDamnedLies, Salvi2025ConversationalPersuasivenessGPT4} -- showing
promise for scalably and durably dissuading people from believing misconceptions such as conspiracy theories \cite{Costello2024DurablyReducingConspiracy} through the fine-tuning of pre-trained models to align with user reasoning and needs \cite{levers}. 
While persuasive LLM outputs are shown to (1) influence public opinions around political topics \cite{levers} (\textit{e.g.,} transgender rights \cite{trans}, climate scepticism \cite{climate}) (2)  help groups find common ground on such subjects~\cite{Tessler2024CommonGround, Argyle2024AIDiscourse, Konya2025UsingCollectiveDialogues}
and (3) reduce beliefs conspiracy theories \cite{Costello2024DurablyReducingConspiracy}, it is unclear how they might affect gig worker awareness and attitudinal change around existing myths. 
To bridge this gap, we recruited crowdworkers from Prolific to explore their beliefs and determine whether a persuasive LLM intervention can provoke critical reflection that shifts their perspectives away from the five gig work myths. 

In Phase~I, we presented five common myths about gig work conditions and asked workers to rate their level of agreement with each. Then, for Phase~II, we invited workers who agreed with at least one myth to write a justification of their position. Through thematic analysis, we uncovered insights into \textit{why} workers agree with these myths, as well as reasons for limited awareness around vulnerabilities and the lack of labor rights.

Beyond initial reactions to the myths, participants who agreed with a myth were presented with either an author-written or LLM-generated rationale to challenge their perceptions while carefully avoiding potential harm. Our experimental setup compared the effects of different persuasion strategies on changes in workers' perceptions. First, since field-specific terminology has been shown to make explanations more satisfying --- laypeople assume that terminology fills gaps in explanations~\cite{Cruz2025HowLaypeopleEvaluate} --- our \textit{Expert Knowledge} condition presented counter-rationales against myths that authors composed based on existing gig work scholarship.
Second, we enhance persuasiveness in the \textit{Selected Expert Knowledge}, \textit{LLM Persuasive} and \textit{Expert Persuasive} conditions by instructing LLMs to personalize to individual workers' written rationales for believing in myths. 
Third, we incorporated both field-specific terminology and LLM-generated content in conditions \textit{Expert Persuasive} and \textit{Combined} since the practice of including external references is shown to reduce overreliance on inaccurate content generated by LLMs while building appropriate reliance for factual outputs ~\cite{Kim2025FosteringAppropriateReliance}. We sought to answer the following questions:

\begin{itemize}
    \item[\textbf{RQ1}:] How do crowdworkers perceive, recognize and understand vulnerabilities in platform-based work? 
    \item[\textbf{RQ2}:] To what extent can these perceptions transform through exposure to persuasive interventions such as LLM-personalised, expert-curated and LLM-generated counter-rationales?
\end{itemize}

Our Phase~I findings show how gig workers’ beliefs in myths are widespread (\S\ref{sec:phase1_quant}). Phase~II results indicate that the underlying reasons for these beliefs are often platform-specific, resulting in varied desires for independence. Furthermore, we demonstrate that exposure to counter-rationales (regardless of condition) can scalably reduce worker agreement with common myths about gig work conditions. We discuss implications of these insights for informing scalable interventions that raise awareness for long-term health and privacy risks in gig work, policies that improve transparency and accountability in platform practices, as well as worker-facing tools that facilitate collective action and worker advocacy.

\section{Related Work}
To investigate the potential for LLMs to reduce myths for platform-based gig workers, we situate our study within extant literature that investigates their work conditions and discuss how persuasive LLM techniques can shift human perspectives and attitudes. \S\ref{myths} further details the literature that informed our expert-curated rationales.

\subsection{Mystified Perceptions of Gig \& Crowd Work Vulnerabilities}
Platform-based gig work offers a wide variety of services, ranging from online crowdwork~\cite{quantifying} and freelancing~\cite{personal} to on-demand ride-hail~\cite{standing} and delivery~\cite{indie} to in-home care~\cite{beyond2} and hospitality~\cite{hospitality}. However, a burgeoning body of work illustrates how algorithmic management strategies~\cite{lee_am, literacy} and information asymmetries~\cite{asymmetry} harm gig worker health, well-being, and welfare by concealing poor work conditions and invisibilizing undocumented labor~\cite{quantifying, invisible, beyond2}. 
But such vulnerabilities and responsibilities remain unobservable to the public~\cite{Hsieh2023GigWorkersWellbeing, HsiehSupporting, Freni-Sterrantino2021PleaNeedInvestigate}, including customers of gig services ~\cite{standing, triangle}, thus impeding progress for civic engagement and regulatory advancement~\cite{Hsieh2023DesigningInterventionsGigWorkers, probes}. Aggravating the situation, workers lack unifying digital organization structures to understand how vulnerabilities materialize, permeate and vary across worker groups \cite{Hsieh2023DesigningInterventionsGigWorkers} and service domains~\cite{ticona2019beyond} --- preventing the formation of collectivism and solidarity~\cite{privacy, flexibility}. 

In fact, prior investigations suggest that crowdwork platforms may intentionally design digital infrastructures that perpetuate competition and isolation while limiting informational and communicative exchanges among workers ~\cite{competition, morality}, giving rise to widespread and inaccurate perceptions (or \textit{myths}) of platforms. Due to such misguided beliefs, workers unintentionally assume various un-negotiated accountabilities from platforms --- \textit{e.g.,} ensuring transactions are smooth and sound, preventing abusive behaviours and arbitration during conflicts ~\cite{unfairness} --- which further compound on already poor work conditions. ~\citet{justice} found that such unrealistic expectations produce perceptions of (organizational) injustice, along dimensions of planning insecurity (or precarity), lack of transparency in performance evaluations, unclear task briefings, and low enumeration -- which correlates with intrinsic crowdworker motivation to complete tasks of varying characteristics~\cite{nature}. \citet{cherry} characterizes platforms' avoidance of such accountabilities as ``crowdwashing'', motivating them to create misnomers (or \textit{myths}) to preserve public image~\cite{rise, networked}. Despite these well-known myths, we are unaware of prior interventions designed to dissuade crowdworkers from believing them.


\subsection{Persuasion Techniques}
Prior work on persuasion and sensemaking indicates that under uncertainty and limited verifiability, people tend to evaluate explanations based on narrative coherence, fluency, and sources rather than factual accuracy~\cite{Petty1986CommunicationPersuasion, Metzger2013CredibilityTrustInformation}. In opaque sociotechnical systems, such as platform-mediated labor, these dynamics make everyday explanations particularly persuasive despite their partial or speculative nature~\cite{Eslami2016FirstItThen, Karizat2021AlgorithmicFolkTheories}. Explanations that are consistently repeated or framed as informative are likely to be adopted and shared, even when their empirical validity is uncertain~\cite{Hasher1977FrequencyConferenceReferential, Fazio2015KnowledgeDoesNot}.

LLMs systematically reproduce these persuasive cues by generating fluent, coherent explanations in conversational form -- often generating confident and persuasive but incorrect responses, which can effectively shape users' reasoning processes ~\cite{Jones2024LiesDamnedLies, Salvi2025ConversationalPersuasivenessGPT4, Goldstein2024HowPersuasiveAIgenerated}. 
\citet{Lu2025HowAISources} found that humans are more open to receiving machine-generated counter-attitudinal messages on controversial issues because they perceive them as less biased, more informative, and having less persuasive intent than human ones. 

Alarmingly, incorrect AI explanations have been shown to convince people even more frequently than correct ones~\cite{Danry2025DeceptiveExplanationsLarge}, while harmful hidden incentives of LLMs influence users' attitudinal shifts more than prosocial ones \cite{shen2026hidden}.
These persuasive effects extend beyond factual reasoning, influencing people's political opinions~\cite{Hackenburg2024EvaluatingPersuasiveInfluence, Argyle2024AIDiscourse, Bai2023ArtificialIntelligenceCan, levers, Lin2025PersuadingVotersUsing} and even durably reducing beliefs in conspiracy theories~\cite{Costello2024DurablyReducingConspiracy}. 
Regardless of whether the messenger is described as human or AI, generated messages are persuasive -- indicating strong persuasive power~\cite{Boissin2025DialoguesLargeLanguage}. In fact, in multi-round debate settings ~\cite{Salvi2025ConversationalPersuasivenessGPT4}, LLMs are shown to be more persuasive than human conversational partners.
However, this persuasive strength of LLMs is coupled with a known tendency to hallucinate inaccurate or fabricated information \cite{costello2026large}, raising concern especially given their ability to scalably personalize to the psychological profile of its users \cite{matz2024potential}. Coupled with their ability to psychologically manipulate human decision-making under financial and emotional contexts \cite{sabour2025human}, LLMs carry risks of scalably producing social harms to users. Despite such known powers and risks of LLM persuasion, we are currently unaware of their capabilities for steering opinions of gig workers towards more accurate realities and factual beliefs.

To mitigate risks for hallucinations to impact people's perceptions, four of our six conditions incorporated expert knowledge around gig work, since including field-specific terminology has been shown to make explanations more satisfying, as readers often assume that the technical language signals depth and accuracy~\cite{Cruz2025HowLaypeopleEvaluate}. 
However, for conditions that combine LLM-generated content with domain-specific terminology, we minimize likelihoods for users to believe in inaccurate but persuasive content by pairing field-specific language with explicit references to accurate sources --- in line with ~\citet{Kim2025FosteringAppropriateReliance}, who found that including sources helps users to distinguish between correct and incorrect answers. 
Inconsistencies hereby serve as reliability cues that allow users better calibrate their trust in the system ~\cite{Kim2025FosteringAppropriateReliance}. 
By using author-written rationales grounded in gig work literature, our design provides source-backed scaffolding that directs persuasive effects toward accurate content instead of well-written but incorrect language.

\subsection{Towards Collectives that Understand \& Contest Gig Work Vulnerabilities}
In place of standard workplace mechanisms for social support, gig workers often turn to online forums or groups to connect with fellow peers~\cite{Holikattiairbnb, social, Yao2021TogetherAlone, quallm}. 
Prior studies uncovered and introduced ways for gig workers to track \cite{accountable}, make sense of ~\cite{Imteyaz2024Gigsense} and share experiences ~\cite{Hsieh2025Gig2GetherDatasharingEmpower}, indicating an intrinsic need to mutually understand work conditions and grievances.
But while such spaces theoretically enable experience-sharing, many workers hesitate to disclose practices, earnings, or complaints in reality, to avoid oversharing with competitors~\cite{Yao2021TogetherAlone} as well as retaliation from platforms. 
Even when relevant knowledge and strategies are publicly documented online, the absence of information organization structures in digital spaces \cite{yao} make it difficult for workers to transmit relevant and timely information across generations.


To better connect the gig workforce and empower its workers in collectively resisting and contesting subpar conditions, its digital communities must first form shared and sustainable cognitive awareness and understandings --- often referred to as algorithmic folk theories \cite{folk}, collective intelligence \cite{CI} or the Mutual Theory of Mind \cite{mutual} --- of how platform algorithms function and embed within their realistic sociomaterial realities.
Given its highly digital nature, \textbf{crowdwork} and its fragmented worker communities particularly exemplify how gig workers are dis-embedded from social protections \cite{networked}. 
In this study, we investigate whether the persuasive power of LLM-generated arguments can challenge workers' preconceived but incorrect understandings of gig work conditions, in pursuit of revealing and charting potentials for LLMs to scalably and durably demystify platform-led misconceptions.

\section{Method}
To investigate worker beliefs in work-related myths and the effect of persuasion techniques for changing their attitudes, we designed a two-phase study in which Prolific participants interact with five common myths about vulnerable gig work conditions. In Phase~I, participants rated their agreement with each `\textit{myth}', which we presented as statements to prevent bias. In Phase~II, we invited people who agreed with at least one of the myths to write a brief response explaining their view. We subsequently presented them with a condition-specific counter-rationale and asked them to re-rate their agreement and provide a new explanation for their response.
The study was conducted through a custom web interface, built using the React framework Next.js\footnote{\url{https://nextjs.org/}}, giving us control over the assignment of myths and rationale conditions.   

\subsection{Myths} \label{myths}
Drawing on gig work literature, two authors constructed five myths to capture a broad range of commonly experienced but underexposed work vulnerabilities, presented below. These were then refined in consultation with the larger research team. 
\begin{center}
    \rule{0.75\linewidth}{0.8pt}
\end{center}
\paragraph{Inequitable Pay (in short: \underline{Equal Pay)}}
Low wages represent a central issue that historically drove workers to protest and unionize for improved treatment. In the case of platform-based gig work, algorithmic wage discrimination~\cite{gender_platform, dubal2023algorithmic, Airbnb} further hinders workers from meeting earning goals with certainty. However, inexperienced workers are unlikely to learn such experiences of financial precarity ahead of time~\cite{finance}, making it crucial to bring the issue to the forefront of public discourse, so as to financially protect younger and novice gig workers. 

\paragraph{Health \& Safety Hazards (\underline{Health \& Safety})}
While the consequences of health and safety hazards are typically not as immediate or omni-present as wage discrimination in gig work, their consequences are often debilitating and compounding. For physical services, workers on the road are constantly under risk of encountering accidents or (sexual) violence~\cite{more, festering}, while caregivers continuously brave the dangers of entering strangers’ homes~\cite{health}. Even for remote (crowd and freelance) work, non-regulated hours and algorithmic management cause higher rates of physical strain~\cite{fatigue}, psychological stress~\cite{longitudinal, hazards} and even morbidity~\cite{nonstandard}. Unlike traditional jobs, gigs imposes such hazards without safety net protections -- \textit{e.g.,} health insurance, workers’ compensation, training, on-site supervision.

\paragraph{Surveillance \& Privacy Gap (\underline{Privacy})} 
Platforms rely on extensive data collection to algorithmically control workers, most of which is not accessible to workers themselves. Such opaque data practices~\cite{accountable, Rao2024Rideshare} prevent workers from collecting evidence around invasive monitoring practices --- further limiting public awareness. Data types collected for physical gig work include telemetry related to movement, location, and performance~\cite{appetite}. Freelancers and crowdworkers experience more fine-grained monitoring through timed screenshots and keystroke captures~\cite{privacy}. Home-based service providers are subject to surveillance through in-home cameras~\cite{watching}. Such constant and intrusive data capture allows algorithms to select projects and clients on the worker’s behalf~\cite{surveillance, monitor}, enacting managerial control that reduces worker agency and privacy.

\paragraph{Illusion of Flexibility (\underline{Flexibility})} Platforms often advertise offerings of ``flexible work arrangements''~\cite{do, hrm}, but their oversaturated labor pool allows them to create highly competitive environments, so much that workers cannot afford to simultaneously earn a living wage and exercise marketed freedoms~\cite{flexibility, standing, running}. 
Meanwhile, financial precarity and low wages (results of high competition) force workers to remain available for late, long and irregular hours~\cite{night, jit, developing}, especially in digitally-mediated crowd and freelance work. In the case of on-demand app workers, algorithmic management mechanisms strip away drivers’ rights to choose times for non-work activities (\textit{e.g.,} holidays, sports events) by limiting the time allotted to accept or decline a gig, forcing workers to forego the most basic necessities such as (bathroom) breaks~\cite{Airbnb}.

\paragraph{Absent Labor Rights (\underline{Labor Rights})} Underlying such absence of worker rights and protections is the fundamental collective bargaining power that gig workers do not possess, due to their legal categorization as independent contractors~\cite{war}. As non-employees, gig workers forfeit numerous protections and benefits such as minimum wage, workers' compensation, health insurance, and nondiscrimination~\cite{hamilton, Hsieh2023DesigningInterventionsGigWorkers}. Moreover, independent contractors are non-exempt from antitrust laws (designed to protect against monopolistic practices), which prevent meaningful forms of unionizing to bargain for the missing but necessary rights and protections~\cite{antitrust, antitrust_exemption}.

\begin{center}
    \rule{0.75\linewidth}{0.8pt}
\end{center}
We presented these five myths to crowdworkers as the following:

\begin{enumerate}
    \item Currently, any two platform-based workers who put in the same amount of labor and hours for the same task are compensated equally.
    \item Gig workers across platforms are currently protected from work-related compromizes to health and safety such as injuries, illnesses, mental health risks, physical strain, etc.
    \item Gig workers across platforms are currently free from monitoring and privacy intrusions during work hours.
    \item Gig workers across platforms currently have flexibility to make a living during hours that are convenient for them.
    \item Gig workers across platforms currently get to take advantage of typical labor rights such as collective bargaining rights, workers’ compensation right, minimum wage, nondiscrimination, etc.
\end{enumerate}

\subsection{Study Procedure}
This study first obtained approval from the Institutional Review Board at Carnegie Mellon University. To start, we presented Prolific participants with instructions and their rights to seek informed consent. Next, participating workers rated their agreement with all myths on a scale from -50 (Strongly Disagree) to 50 (Strongly Agree), with 0 being neutral. We chose this scale to capture both belief direction (+/-) and magnitude, allowing for a more granular assessment of belief strengths and attitudinal changes. Myths were randomized to prevent ordering effects.

To assess the ability of our intervention to challenge the opinions of workers who believed in these myths, we recruited only participants who agreed with at least one myth in Phase~I for Phase~II. We asked these workers to (1) write a response justifying their position for a myth they believed, (2) provide an example from their experience to support their views, and (3) rate their own confidence in the response on a 0 to 100 scale to capture fine-grained differences.
Among the myths each participant agreed with, we selected the least-represented one in the overall data set to balance response rates across myths. 

After their initial responses, we presented participants with an author- or LLM-written counter-rationale elaborating on the vulnerabilities associated with the selected myth. Participants were subsequently asked to (1) rate their agreement with the myth again and (2) write a brief response describing what led them to change or maintain their level of agreement. As these counter-rationales were produced in response to workers' own rationales for endorsing myths, their content systematically challenged those misbeliefs, with the intention of non-maliciously influencing worker opinions. Our aim was for workers to learn from the rationales while minimizing risks of harm. To discourage crowdworkers from using generative AI to compose their written responses, 
we disabled pasting on the task page and
warned them about potential rejections of their submission. Participants were able to opt out of Phase~II and end the study.

We recruited participants on Prolific, offering compensation of \$2 for Phase~I and a bonus of \$3.75 when they are selected for Phase~II, with both compensations corresponding with an hourly payment of \$15. 
Since the myths address U.S.-specific vulnerabilities (which can vary substantially across countries and regions), we recruited participants only from the United States. 
As crowdworkers, our Prolific participants allowed us to study vulnerability perceptions for one subset of gig workers at scale. We intentionally employed a platform-agnostic framing when composing the myths to broaden their applicability to gig platforms beyond crowdwork, thereby inviting more pluralistic perspectives from participants who also work on other platforms.

During piloting, we analysed participants' responses and identified misinterpretations of the myths, iteratively refining them to improve clarity. To further reduce ambiguities in our terminology, we added a definition of gig workers and platforms to an information modal for participants to consult. We also added the term `currently' to emphasize each myth's purpose of representing present conditions rather than idealized situations that \textit{should} be pursued. In the piloting phase, participants consistently found the myths relevant to their work, confirming our expectation that the platform-agnostic myths would resonate with their experiences.
\subsubsection{Measures}
For the two phases, we collected the following:
\begin{itemize}
    \item \textbf{Initial agreement}: Participants scored their agreement with all five myths in Phase~I on a scale from -50 (Strongly Disagree) to 50 (Strongly Agree), with 0 being neutral. This bipolar scale was chosen to reduce potential confusion when interpreting midpoints (\textit{e.g.,} 50 on a 0–100 scale).
    \item \textbf{Participant response}: When selected for Phase~II, participants wrote a response (of at least 100 words) to justify their score on a myth they agreed with.
    \item \textbf{Participant example}: Participant responses are accompanied by an example (of at least 75 words) to illustrate their view based on experience.
    \item \textbf{Participant confidence}: Participants indicated self-confidence (0 to 100) in their written response, using a unipolar scale appropriate for measuring the degree of certainty.
    \item \textbf{Final agreement}: Participants reconsidered their agreement with the selected myth in Phase~II --- on a scale from -50 (Strongly Disagree) to 50 (Strongly Agree) --- after reading the rationale we presented them with.
    \item \textbf{Final participant response}: Participants wrote a response (of at least 100 words) to explain their final agreement score after reading the condition-dependent rationale we presented.
\end{itemize}

After the study activities described above, we collected demographics and post-measures. For all participants, we recorded their age, gender, education, how long they had worked as a gig worker, the amount of reliance on gig work for income, and whether they had a traditional job. Using a 7-point Likert scale (to minimize cognitive load while maintaining reliability), we also asked workers if they related to the presented topics, whether the topics were relevant to their work situation, and how important they perceived it to be for gig workers to collectively bargain for policy improvements. Phase II participants were also asked to rate (on a 7-point Likert scale) and elaborate on how well they related to the presented rationale, how much they learned from the myths, and whether the tone of the presented rationale was appropriate.

\subsubsection{Counter-rationale Conditions}
Phase~II Participants were randomly assigned to one of  six conditions, which determined the style of the counter-rationale they received after providing their initial response:

\begin{enumerate}
    \item \textit{Expert Knowledge}: Static rationale written by the authors based on scientific sources (see Appendix \ref{app:author_rationales} for the full author-written rationales).
    \item \textit{Selected Expert Knowledge}: Prompts an LLM to select the arguments from the author-written rationale that are most relevant to the participant's written response.
    \item \textit{LLM Baseline}: Prompts an LLM to respond to the participant's written justification.
    \item \textit{LLM Persuasive}: Builds on the \textit{LLM Baseline} condition but adds instructions to the LLM prompt to phrase the response in a persuasive manner.
    \item \textit{Expert Persuasive}: Builds on the \textit{Selected Expert Knowledge} condition but additionally instructs the LLM to phrase it in a persuasive manner.
    \item \textit{Combined}: This condition instructs the LLM to combine the expert knowledge from the author-written counter-rationale with additional arguments, and incorporates both the LLM selection and persuasive layers.
\end{enumerate}
\begin{table*}[]
\centering
\caption{Participant engagement with platforms other than Prolific}
\resizebox{\textwidth}{!}{%
\begin{tabular}{lrrrrrrrrrrrr}
\toprule
\textbf{Platform} 
& \textbf{Any other platform} 
& \textbf{MTurk} 
& \textbf{DoorDash} 
& \textbf{Upwork} 
& \textbf{Uber} 
& \textbf{Fiverr} 
& \textbf{Freelancer} 
& \textbf{TaskRabbit} 
& \textbf{Rover} 
& \textbf{Lyft} 
& \textbf{Grubhub} 
& \textbf{Other} \\
\midrule
\textbf{Participants (\%)} 
& 58.50 
& 24.57 
& 9.75 
& 9.32 
& 8.05 
& 5.51 
& 2.54 
& 2.12 
& 1.69 
& 0.85 
& 0.42 
& 20.80 \\
\bottomrule
\end{tabular}}
\label{table:platform_activity}
\end{table*}

\begin{figure}[h!]
  \centering
  \includegraphics[width=\linewidth]{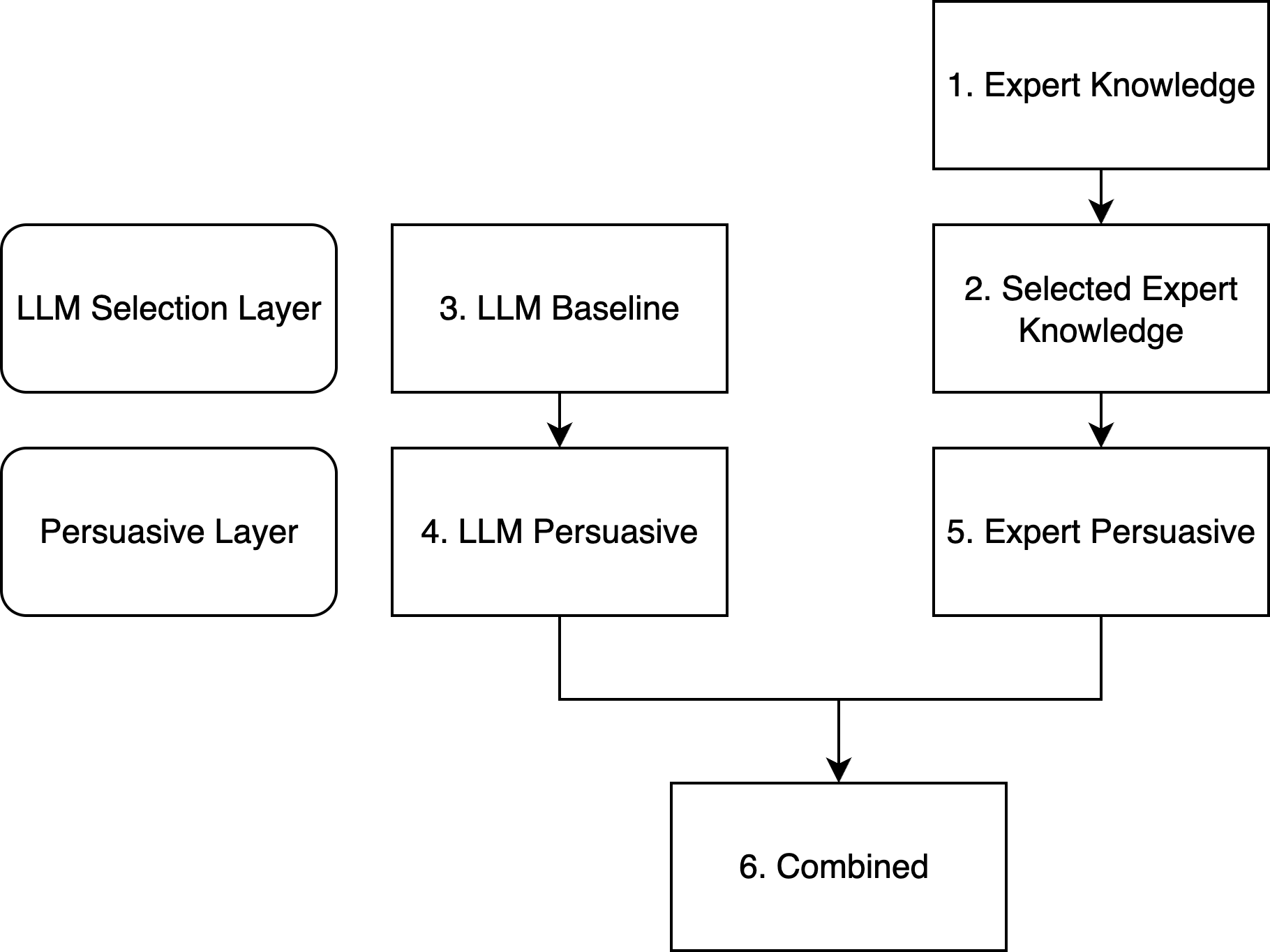}
  \Description{Flowchart of the layered conditions, adding the LLM Selection Layer to the LLM Baseline on the left-hand side and the LLM Persuasive layer underneath. The right-hand side starts with Expert Knowledge, adds the LLM Selection Layer in Selected Expert Knowledge, and the Persuasive Layer in Expert Persuasive. Arrows from both sides point to the Combined condition. }
  \caption{A flowchart of the layered condition design, adding LLM selection of arguments to best match the worker's response in the \textit{LLM Selection Layer} and instructions to enhance persuasiveness in the \textit{Persuasive Layer}. The \textit{Combined} condition instructs the LLM to combine expert knowledge with additional arguments, and additionally incorporates both the LLM selection and persuasive layers. Except those in the \textit{Expert Knowledge} condition, all counter-rationales were LLM-generated.}
  \label{fig:conditions_flowchart}
\end{figure}

\paragraph{Layered Conditions}
The conditions followed a layered design, with expert knowledge (top), LLM selection (middle), and persuasive prompting (bottom) added at each layer. Figure \ref{fig:conditions_flowchart} provides an overview of the layered rationale conditions. The LLM-generated conditions (all except the static \textit{Expert Knowledge} of the top layer) were generated live after participants wrote their responses, using GPT-4o (model: gpt-4o-2024-05-13).

\paragraph{Top layer: \textbf{Expert Knowledge}} Author-written counter-rationales that provide evidence against the myths based on gig work literature. The \textit{Expert Knowledge} condition presents this text verbatim, while the \textit{Selected Expert}, \textit{Expert Persuasive}, and \textit{Combined} conditions use it as an input prompt.

\paragraph{Middle layer: \textbf{LLM Selection}} All LLM counter-rationale conditions are personalized to participants' written responses.

\paragraph{Bottom layer: \textbf{Persuasive LLM Prompt}} In the persuasion layer, prompt adaptations (based on~\citet{Costello2024DurablyReducingConspiracy}) are used to instruct the LLM to use all persuasive tools to its disposal to formulate a convincing counter-rationale. See Appendix~\ref{app:full_prompts} for full prompts and Appendix~\ref{app:generated_rationales} for an example of generated rationales for each rationale condition. 

\subsubsection{Thematic Analysis}
First, we conducted a focused thematic analysis~\cite{braun2006} of participants' initial written responses to explore the reasons behind their agreement ratings and understand why they believed in the myths. Two authors independently familiarized themselves with the data and conducted open coding of the responses to gain an initial understanding and to identify and discuss potential themes. 

Building on this, the first author conducted a fine-grained analysis of all written responses using the developed coding scheme, assigning codes to recurring ideas within each myth. These codes were iteratively grouped into broader themes across the dataset. The resulting themes and their interpretations were subsequently reviewed and refined in discussion with the research team to ensure consistency and shared understanding. 

The analysis of the reconsidered responses followed a similar approach, but adopted a more comprehensive lens to explore (1) initial responses written by participants, (2) presented counter-rationales, and (3) reconsidered participant responses together -- relating these to participants’ quantitative agreement scores to understand what led to changes or maintenance of their stance.

\subsection{Positionality}
The authors are an international team based at two universities. Two authors have extensive experience in the gig work context, and have previously worked to center gig worker well-being in their research. They are also aware of how academic and media narratives can pressure us to emphasize and prioritize negative harmful consequences of platform-based labor on worker wellbeing. We have approached this study with the intent to balance such narratives while focusing on worker well-being. Our aims as researchers are to debunk what we conceive as myths in the gig work context, while recognizing that these may not reflect every worker's lived experience. While we put well-being center, we take care to reflect throughout the study to avoid adopting a savior complex in our academic inquiry. We remain cautious about assuming that techno-solutionist approaches can adequately address these vulnerabilities, and recognize that their uncritical use could risk exacerbating them. 

In particular, the persuasive elements of our study design can influence worker behaviors in ways that produce financial and material consequences. As a result, we also reflect on how our objective of changing workers’ perceptions may affect their intrinsic motivations to engage in platform work, and consequently their financial stability, while being mindful not to discourage participants who rely on gig work as their primary source of income. Finally, while our intention to provoke worker reflection of their own work conditions are meant to encourage stronger labor protections and healthier work practices, we recognize that diverse and pluralistic nature of individual worker experiences and take care to minimize potentials of trivializing workers' own initial perceptions or delegitimizing their lived experiences.

\section{Results}

\subsection{Phase~I: Demographics \& Initial Agreement with Myths} \label{sec:phase1_quant}
 We recruited 236 participants: 121 were female, 110 male, 1 agender, 1 non-binary, 1 transgender, and 2 undisclosed. The average age was 39.93 years ($SD = 11.92$, ranging from 19 to 72), while the highest level of completed education varied: Primary school (0.42\%), Secondary school (29.24\%), Bachelor's degree (47.46\%), Master's degree (18.64\%), and doctorate (4.24\%). Participants' gig work experiences ranged from less than 3 months (2.97\%), 3-6 months (4.66\%), 6-12 months (16.95\%), 1-2 years (23.31\%), 2-5 years (24.15\%), to 5 or more years (27.97\%). Income-wise, 43\% considered gig work an essential source, while 49\% did not, and 7.62\% preferred not to disclose their income reliance. Occupation-wise, 49.57\% held another full-time job, 24.15\% had another part-time job, 24.58\% had no other job, and 1.69\% did not disclose their job status. Apart from Prolific, a majority of participants (\textbf{58.50}\%) were active on other platforms, as shown in Table \ref{table:platform_activity}.
 

Participants found the topics discussed in the myths to be relevant to their work ($M = 5.39$, $SD = 1.87$) and also considered it important for gig workers to engage in collective bargaining ($M = 5.48$, $SD = 1.79$). Both measures were rated on a 7-point Likert scale. Table~\ref{table:statement_responses} shows the means of workers' agreement values per myth during Phase I: most agreed with the \underline{\emph{Flexibility}} myth while \underline{\emph{Privacy}} and \underline{\emph{Equal Pay}} showed mixed responses. Lastly, the myths about \underline{\emph{Health and Safety}} and \underline{\emph{Labor Rights}} showed more disagreement than agreement.

\begin{table}[!h]
\centering
\caption{Counts of participants who agreed (agreement $\geq~0$) or disagreed (agreement $<~0$) per myth. Ratings are reported as mean and standard deviation (SD) from -50 (Strongly Disagree) to 50 (Strongly Agree), with 0 representing neutral.}
\begin{tabular}{lrrcc}
\toprule
\textbf{Myth} & \textbf{Agree} & \textbf{Disagree} & \multicolumn{2}{c}{\textbf{Agreement}} \\
\cmidrule(lr){4-5}
& & & Mean & SD \\
\midrule
Privacy          & 136 & 100 & -0.41  & 27.64 \\
Flexibility      & 220 &  16 & 28.19  & 21.42 \\
Equal Pay        & 103 & 133 & -9.37  & 29.25 \\
Health and Safety&  74 & 162 & -19.26 & 27.12 \\
Labor Rights    &  84 & 152 & -16.67 & 31.24 \\
\addlinespace
All              & 617 & 563 & -3.50  & 17.48 \\
\bottomrule
\end{tabular}
\label{table:statement_responses}
\end{table}

Agreement was especially strong for the myths concerning \underline{\emph{Privacy}} and \underline{\emph{Flexibility}}. Even the least accepted myths, related to \underline{\emph{Health and Safety}} and \underline{\emph{Labor Rights}}, still received agreement from at least one third of participants. Overall, these findings indicate a lack of awareness regarding worker vulnerabilities. Most participants endorsed multiple myths (Table \ref{table:agreement_distribution} shows the distribution for the number of myths believed), confirming that these are widespread among the sample. We examine the vulnerabilities in more depth in Phase~II.

\begin{table}[h!]
\centering
\caption{Distribution of how many myths participants agreed with (by rating agreement $\geq$ 0)}
\begin{tabular}{lrrrrrr}
\toprule
\textbf{Myths Agreed With} & 0 & 1 & 2 & 3 & 4 & 5 \\
\midrule
\textbf{Frequency} & 9 & 43 & 70 & 54 & 28 & 32 \\
\bottomrule
\end{tabular}
\label{table:agreement_distribution}
\end{table}

\subsection{Phase~II: Belief Changes After Exposure to Counter-Rationales}
\paragraph{Selective Sampling}
We invited participants who agreed with at least one myth to take part in Phase~II. To determine the study's required power for the intended model, we conducted an analysis using G*Power~\cite{Faul2007GPower} selecting medium-to-large effect sizes ($f^{2} = 0.15$), an alpha level of 0.05, and a power of 0.8 for the 35 intended predictors in the model, resulting in a sample size of 201. To evenly spread participants across the six conditions (34 each), we recruited 204 participants in total for Phase~II. Figure \ref{tab:participants_count} presents the distribution of participants across myths, which was not perfectly even due to differences in agreement across the myths and the simultaneous nature of task completion on Prolific.

Participants were highly confident in their written responses (\textit{M} = 81.80, \textit{SD} = 19.64; median = 85, \textit{IQR} = 75--100), suggesting that they felt knowledgeable about the presented myths. Table \ref{tab:post_measures} shows the specific post measures for Phase~II participants, all rated on a 7-point scale. The \textit{Expert Knowledge} and \textit{Expert Persuasive} conditions were perceived as less appropriate in tone and participants related less to these rationales. Participants learned the most from the \textit{LLM Persuasive} condition and least from the \textit{Expert Persuasive} one. However, these differences were not significant when tested through Tukey-adjusted pairwise comparisons.

\begin{table}[!h]
\centering
\caption{Participants assigned to each myth in Phase~II.}
\label{tab:participants_count}
\begin{tabularx}{\columnwidth}{XXXXX}
\toprule
\textbf{Privacy} & \textbf{Flexibility} & \textbf{Equal Pay} & \textbf{Health and Safety} & \textbf{Labor Rights} \\
\midrule
49 & 47 & 34 & 36 & 38 \\
\bottomrule
\end{tabularx}
\end{table}

\begin{table*}[!h]
\centering
\caption{Participant ratings for the appropriateness of the counter-rationale's tone, how much they learned about the vulnerabilities, and how much they relate with the counter-rationale, across conditions and overall.}
\centering
\begin{tabular}[t]{lcccccc}
\toprule
\multicolumn{1}{c}{ } & \multicolumn{2}{c}{\textbf{Tone}} & \multicolumn{2}{c}{\textbf{Learned about vulnerabilities}} & \multicolumn{2}{c}{ \textbf{Relate with Counter-rationale}} \\
\cmidrule(l{3pt}r{3pt}){2-3} \cmidrule(l{3pt}r{3pt}){4-5} \cmidrule(l{3pt}r{3pt}){6-7}
 \textbf{Counter-rationale Condition} & Mean & SD & Mean & SD & Mean & SD\\
\midrule
Expert Knowledge & 5.57 & 1.67 & 5.03 & 1.67 & 4.80 & 1.88\\
Selected Expert Knowledge & 6.27 & 0.84 & 4.97 & 1.93 & 5.33 & 1.74\\
LLM Baseline & 6.32 & 1.25 & 5.06 & 2.06 & 5.21 & 1.86\\
LLM Persuasive & 6.18 & 1.11 & 5.35 & 1.43 & 5.15 & 1.96\\
Expert Persuasive & 5.56 & 1.94 & 4.82 & 1.85 & 4.38 & 2.20\\
Combined & 6.24 & 1.16 & 5.12 & 1.53 & 5.35 & 1.57\\
\addlinespace\\
All & 6.02 & 1.40 & 5.06 & 1.74 & 5.03 & 1.89\\
\bottomrule
\end{tabular}
\label{tab:post_measures}
\end{table*}

\subsubsection{Agreement Change} \label{sec:agreement_change}
\begin{figure*}
    \centering
    \includegraphics[width=\textwidth]{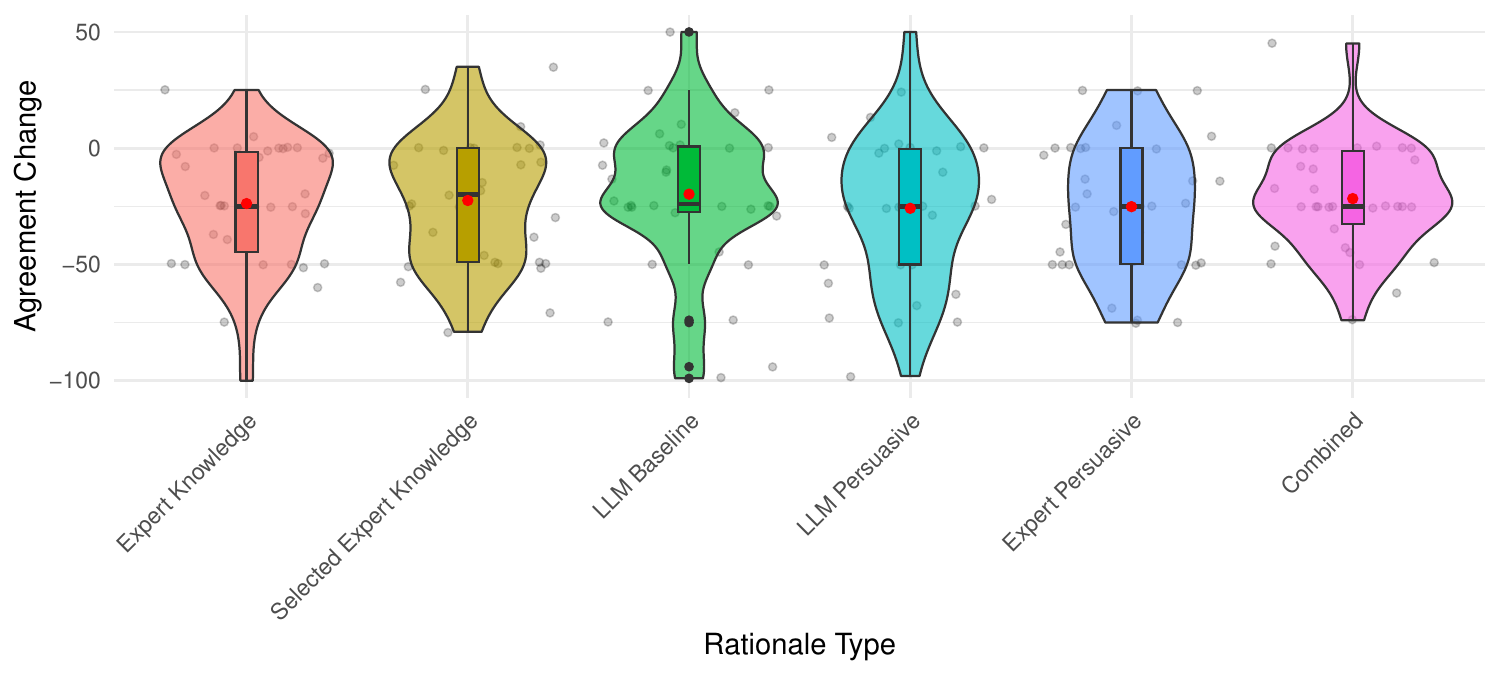}
    \Description{Violin plots of the agreement change per condition, showing that there are no significant differences between the conditions.}
    \caption{Agreement change (agreement in Phase~II - agreement in Phase~I), outlined per condition. }
    \label{fig:agreement_change_condition}
\end{figure*}

\begin{figure*}
    \centering
    \includegraphics[width=\textwidth]{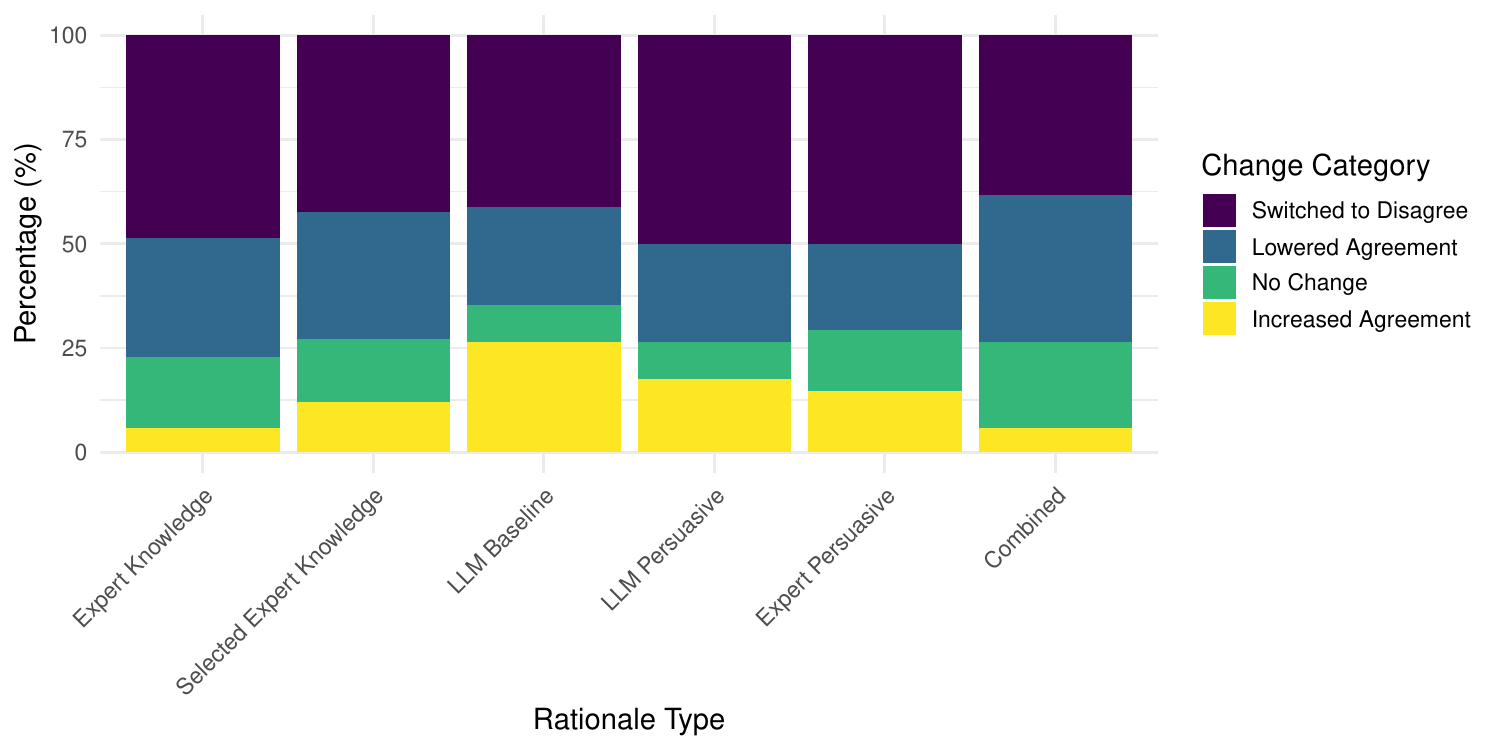}
    \Description{Bar charts of the percentage of participants who changed their agreement per condition. Expert knowledge, LLM Persuasive, and Expert Persuasive switched to disagreeing with the myths the most, at just under 50 percent. The other conditions also show high switch rates. For most conditions, 75 percent of participant either switched to disagreeing with the myth or lowered their agreement. }
    \caption{\textbf{Switch rate by condition}. Switched to disagree means that people agreed in Phase~I but disagreed in Phase~II. Lowered agreement means they lowered their final agreement in Phase~I but did not switch to a negative agreement. }
    \label{fig:switch_rate}
\end{figure*}

Participants showed a substantial overall agreement change ($M = -23.13$, $SD = 28.36$), on a scale from -50 (Strongly Disagree) to 50 (Strongly Agree), with 0 being neutral. As illustrated in Figure \ref{fig:agreement_change_condition}, agreement change scores were similar across conditions. After reading the rationale, 92 participants (45\%) shifted to disagreement in Phase~II, 55 (27\%) lowered their agreement, 28 (14\%) increased their agreement, while 29 (14\%) did not change. Figure \ref{fig:switch_rate} presents the switch rate per condition.

\subsubsection{Model Evaluation}
To better understand agreement changes, we model the data through a linear model predicting participants' final agreement. We started our model construction with 35 possible predictors and conducted model selection through incremental removal of variables based on their predictive power and Akaike information criterion (AIC)~\cite{Akaike1974NewLookStatistical}. The main removal is the interaction effect between condition and myth, which did not yield any significant results. This removal resulted in a model with a lower AIC score and thus better fit. Our final model is the following:

\begin{align*}
\text{Final Agreement} \sim\ & \text{Initial Agreement} + \text{Rationale Condition} + {}\\
& \text{Myth} + \text{Gender} + \text{User Confidence} + {}\\
& \text{Rationale Tone} + \text{Learned from Rationale} + {} \\
& \text{Relate with Rationale} + \text{Myth Relevance} + {} \\
& \text{Collective Bargain}
\end{align*}

\begin{table*}[h]
    \centering
    \caption{Results for the linear model predicting participants' final agreement in Phase~II. The model uses `\textit{Expert Knowledge}' as reference level for Rationale Condition, Privacy for Myth, and Female for Gender.}
    \begin{tabular}{@{}lrrrrr@{}}
        \toprule
        \textbf{Predictor} & \textbf{Estimate} & \textbf{Std. Error} &  \textbf{$t$-ratio} & \textbf{\textit{p}-value} & \\
        \midrule
        Initial agreement & 0.48 & 0.11 & 4.53 & $<0.001$ & *** \\
        \addlinespace
        Rationale Condition \\
        \hspace{3mm}Selected Expert Knowledge & -0.04 & 6.02 & -0.01 & 0.995 & \\
        \hspace{3mm}LLM Baseline & 2.69 & 5.99 & 0.45  & 0.654 & \\
        \hspace{3mm}LLM Persuasive & -0.56 & 5.99 & -0.09 & 0.925 & \\
        \hspace{3mm}Expert Persuasive & -4.69 & 5.90 & -0.80 & 0.427 & \\
        \hspace{3mm}Combined  & 3.23 & 5.99 & 0.54  & 0.591 & \\
        \addlinespace
        Myth \\
        \hspace{3mm}Flexibility & 19.64 & 5.51 & 3.56 & $<0.001$ & *** \\
        \hspace{3mm}Equal Pay & 0.12  & 5.54 & 0.02 & 0.982 & \\
        \hspace{3mm}Health and Safety & -2.01 & 5.46 & -0.37 & 0.714 & \\
        \hspace{3mm}Labor Rights & -1.18 & 5.35 & -0.22 & 0.826 & \\
        \addlinespace
        Gender \\
        \hspace{3mm}Male & 6.99 & 3.47 & 2.02 & 0.045 & {*} \\
        \hspace{3mm}Prefer not to say & -12.45 & 18.14 & -0.69 & 0.493 & \\
        \hspace{3mm}Prefer to self-describe & -19.92 & 17.53 & -1.14 & 0.257 & \\
        \addlinespace
        User confidence & 0.15 & 0.09 & 1.64 & 0.102 & \\
        Rationale tone & -0.35 & 1.42 & -0.25 & 0.803 & \\
        Learned from rationale & -3.70 & 1.11 & -3.34 & 0.001 & ** \\
        Relate with rationale & -0.43 & 1.17 & -0.36 & 0.716 & \\
        Myth Relevance  & 2.54 & 1.05 & 2.41 & 0.017 & {*} \\
        Need for collective bargaining & -3.58 & 1.08 & -3.32 & 0.001 & ** \\
        \bottomrule
    \end{tabular}
    \begin{flushleft}\centering{\scriptsize $^{***}p<0.001$;\ $^{**}p<0.01$;\ $^{*}p<0.05$}\end{flushleft}
    \label{tab:model_results}
\end{table*}

A comparison with the null model using an F-test showed that the full model provides a significantly better fit ($F(19, 184) = 7.64$, $p~<~.001$), explaining 44.1\% of the variance in final agreement ($R^2 = 0.44$, adjusted $R^2 = 0.38$).  We assessed the multicollinearity between the model's parameters and found variance inflation factors (VIF) ranging from 0.02 to 1.32, which is well below the commonly used multicollinearity threshold of five to ten~\cite{Hair1995MultivariateDataAnalysis}.

The overview of the model results is presented in Table~\ref{tab:model_results}. Participants who initially agreed more strongly also tended to maintain higher final agreement. Participants who believed they learned from the counter-rationale or perceived the myth as relevant showed lower final agreement, leading to more disagreement with the myth. Similarly, recognizing the need for collective bargaining predicted significantly lower agreement.

Compared to \underline{Privacy}, the model showed significantly higher agreement for the \underline{Flexibility} myth ($\beta = 19.64$, $SE = 5.51$, $t(184) = 3.56$, $p = 0.004$). Tukey-adjusted pairwise comparisons confirmed this pattern: \underline{Flexibility} yielded significantly higher agreement scores than \underline{Privacy} ($\beta = 19.64$, $SE = 5.51$, $p = 0.004$), \underline{Equal Pay} ($\beta = 19.52$, $SE = 5.78$, $p = 0.008$), \underline{Health and Safety} ($\beta = 21.65$, $SE = 5.76$, $p = 0.002$), and \underline{Labor Rights} ($\beta = 20.82$, $SE = 5.65$, $p = 0.003$). No other pairwise differences were significant (all $p > 0.99$).

The difference between conditions were not significant (all $p > 0.43$), confirming that counter-rationale condition do not affect agreement scores. Model results indicate that male participants reported slightly higher agreement than females. Tukey-adjusted pairwise comparisons did not confirm differences between genders (all $p > 0.19$).


\subsection{Initial Written Responses: Motivations for \& Risks of Believing Myths}
Next, we present the results of our thematic analysis. First, we describe why people agreed with the myths, including their reasons for continuing to engage in gig work (\S\ref{sec:phase1_motivations}), which include independence and trust in platforms, as well as recognition of potential risks (\S\ref{sec:phase1_risks}). In \S\ref{sec:qual_phase2} we present our analysis of their written responses to our counter-rationales. The dataset of all participants' written responses is included in the supplementary material, which has been verified to contain no identifiable information.

\subsubsection{Motivations to Engage: Independence \& Flexibility} \label{sec:phase1_motivations}
Participants often explained their own motivations for doing gig work as an explanation for their agreement with a myth. Several shared desires to maintain their independence as a motivating factor: ``\textit{Most of us, absolutely do not want to be considered an employee.}'' (P192).
One underlying reason is a desire for freedom to reject tasks that carry health and safety risks, or occur at inconvenient times: ``\textit{Gig workers can pick and choose which job we want to take on. So if we do not like the pay of one task, we can choose to either skip it and move on, or save it for another time}'' (P181).
This strong perception of work flexibility explains the nearly unanimous agreement with the myth on \underline{\emph{Flexibility}} (see Table \ref{table:statement_responses}).

\paragraph{Trust in Platforms}
In addition to flexibility, participants also expected platforms to address work vulnerabilities. For example, many refused to acknowledge payment disparities among workers and firmly believed myths around pay equity: ``\textit{Any platform-based worker will be paid the same amount. [\dots] This engenders confidence in the platform.}'' (P99).
Moreover, workers trusted crowdwork platforms like Prolific to protect their \underline{\emph{Privacy}},
explaining the higher number of them agreeing with the myth (see Table \ref{table:statement_responses}): 
``\textit{My platform takes extreme measures to protect the privacy of my information and informs [\dots] [requestors] are not permitted to request or ask for personal information prior to beginning the study and throughout the study.}'' (P36).
In contrast, privacy violations by ride-hailing platforms were cited as reasons for distrust. 

Several participants expressed feelings of being protected by \underline{\emph{Labor Rights}} (but could not articulate specific protective mechanisms) while most are aware of this vulnerability, as indicated by the higher rate of disagreement. 
However, relative policy developments in specific states led many workers to believe they are protected, despite a lack of understanding of the exact rights, which also vary
``\textit{depending on the country, the platform and the type of work. Some platforms have introduced basic safety policies but many still fall short.}'' (P166).

\subsubsection{Acknowledging Inherent Risks} \label{sec:phase1_risks}
Despite their trust in platforms, participants also recognized inherent risks, including the lack of choice for vulnerable groups to refuse jobs: ``\textit{Individuals disadvantaged with a lower skillset, lack of education, disabilities, etc. are more likely to be compliant with these unfair labor practices. This can be due to desperation or even lack of knowledge regarding rights.}'' (P196). Despite relatively high proportions of believers in the \underline{\emph{Flexibility}} myth,
some recognized factors such as the need to book time slots for work that make their hours comparable to a regular job, and surge hours that foster the `fear of missing out' (P62). P69 relates how ``\textit{work availab[ility] is extremely variable [\dots] gig work is still beholden to a `normal' schedule more so than many people think.}''



Many workers believed that working at home shields them from physical vulnerabilities such as  dangerous encounters with strangers. P135 shared how ``\textit{At home gig workers don't have to worry about being outside in bad weather, increased car accident risks, physical strain from lifting heavy items.}'' (P135). 
However, P139 acknowledged challenges to working from home, including work-life balance and mental health: ``\textit{You have to balance both your gig work, as well as managing your personal life at the same time.}'' P138 brought up mentally harmful content in their experience:
``\textit{Some of the studies contain sensitive content about abuse, destruction and traumatic injuries. The nature of these studies makes them mentally challenging to deal with. I often think about them afterwards.}''

\subsection{Reconsidered Responses: Worker Reflections on Independence, Trust in Platforms \& Desire for Transparency} \label{sec:qual_phase2}
Our counter-rationales were generally effective at weakening workers' beliefs in prevailing myths, but many participants still resisted these interpretations and refused to acknowledge them to the full extent.
\S\ref{persuasive} shows how factual knowledge and the perceived authority of references critically shape worker understandings and reflection --
supporting the quantitative finding that exposure to counter-rationales reduces beliefs in myths (\S\ref{sec:agreement_change}). 
Many workers who reduced their agreement in myths began demanding further transparency from platforms regarding existing practices, while also seeking strategies to resist them (\S\ref{sec:transparency}).
One dominant motivation for workers to resist our counter-rationales was their underlying desires to preserve independence (\S\ref{independence}).
Meanwhile, \S\ref{sec:persist} reveals the nuanced duality between trust in platforms to uphold worker rights and confidence in online social mechanisms for holding platforms accountable, highlighting workers' gaps in legal knowledge of specific protections and the need for future digital interventions to go beyond public forums and build spaces that foster and sustain worker solidarity.

\subsubsection{Counter-Rationales as Persuasive Authority} \label{persuasive}
Participants generally perceived the counter-rationales as a persuasive authority, which was impactful in lowering their agreement with the myths. While most did not explicitly reference the counter-rationale,  a few
mentioned it as an ``article''. 
Several participants attributed it to a person, researcher, or expert, while a few explicitly called them LLM-generated.
The counter-rationales influenced participants by providing new factual information, reinforcing existing knowledge, prompting reflection on their opinions, or connecting to other articles they had read. 
P44 described the rationale's strong effect on their attitude change:
``\textit{The info presented in the article was what changed my mind  [\dots] I am glad to have read the information. I feel bad for the people who have to work in homes and are monitored.}'' 

Participants considered counter-rationales well-articulated, with the references enhancing their perceived authority by providing new, eye-opening information from credible sources:
``\textit{
After reading the document, it was obvious I was wrong. [...] The document was an eye-opener. Quoting studies would let me check the sources and raises my confidence in the document's conclusions. [...] I am surprised at the result, obviously, since it changed my opinion so much.}'' (P123).

\subsubsection{Need for Transparency \& Resistance Tactics} \label{sec:transparency}
After reading counter-rationales, participants expressed a desire for more platform transparency surrounding the vulnerabilities. For example, concerning \underline{\emph{Privacy}}, those who lowered their agreement shared uncertainty on what data is collected through their platforms and how it is processed:
``\textit{once the work is done, is [the data] being stored, sold to third parties, or is it being deleted? It is important that these platforms clarify exactly what is being monitored, what is being collected, for what purpose and what is being done with that data afterwards.}'' (P18).
The ways that algorithms are used to determine the availability and payment of jobs also mattered. For the \underline{\emph{Equal Pay}} myth, P100 commented how ``\textit{I have no idea what algorithms are being used or not. They might pay old people less, or young people less. They might choose to pay less late at night or early in the morning.} ''

Due to opaque and unregulated platform data practices, alternative communication media such as Reddit became valuable spaces for workers to learn from each other's gig work experiences. P106 shared how online social fora
helped them gain better understandings of \underline{\emph{Fair Pay}} and learn practices such as \textbf{refusal} of lower pay orders:
``\textit{Low paying orders are appealing until you know what you're going to be offered, and that you can refuse orders. However, not everyone knows that, and often will take \$2.25 per order until someone tells them to wait for larger ones. I myself didn't know until I went on the r/DoordashDrivers Reddit page.}'' 

\subsubsection{Worker Independence as an Interpretive Lens} \label{independence}
Participants cited worker independence as the primary reason for doing gig work (see \S\ref{sec:phase1_motivations}). 
However, the counter-rationales made therm recognize that this independence also entails greater exposure to fluctuations in work availability, depending on factors such as region, surge hours, demographics, and worker experience. 
Across myths, participants consistently drew on notions of worker independence as a lens through which they engaged with counter-rationales, enabling acceptance for some while constraining or rejecting broader implications for others. 
This dynamic is evident in participants' reflections on \underline{\emph{Flexibility}}, where some considered how platform incentives shape when work is available and economically viable, and how algorithmic demand and competition compel them to adjust work practices.
These mechanisms made several participants realize the negative impacts of variable work availability on their work-life balance:
``\textit{Choosing to not work during hours that are provide comes at a cost [\dots] So the gig work becomes a bit of a catch 22. If you make yourself endlessly available [\dots] you don't have the benefits an employee has. If you work only on your own terms, you may not make enough money to be comfortable in life.}'' 

However, some workers remained unshifted in their position regarding (schedule) independence, or \emph{\underline{Flexibility}}. For instance, P52 insisted that ``\textit{the OPTION to work at whatever time works for the worker is used by many people. So, gig work [\dots] can provide control over work hours.}''
Others wrote similar reasoning to rationalize unchanged opinions around \underline{\emph{Health and Safety}}, mentioning alternative options to reject stressful jobs, or quit the gig economy altogether. On \underline{\emph{Equal Pay}} and \underline{\emph{Privacy}}, participants similarly reasoned that workers can reject jobs with low payment or ones that infringe on their privacy. 
Even those who began to acknowledge the vulnerabilities still considered it the workers' responsibility to reject the particular jobs that expose them to such risks (P144, P15).



Independence was also the underlying reason for workers' concerns around \underline{\emph{Labor Rights}}, 
since they viewed improved labor rights as incompatible with their independent contractor status. As P192 was adamant to share:
``\textit{Most of us WANT to be independent contractors, and NOT classified as a worker. We do not have any problems with getting to choose what work we do, who we do it for, and when we do it. As I said all my life, `I need no master.'}'' 


\subsubsection{Persisted but Uncertain Trust in Platforms} \label{sec:persist}
Many participants continued to trust their platforms to provide worker protections even after reading counter-rationales that detail how platforms expose them to vulnerabilities, demonstrating deep-rooted trust. 
For instance, workers like P169 still feel certain that their platform offers worker protection, but cannot articulate exactly what these are:
``\textit{Gig workers may be able to take advantage of these labor rights but I am not sure which ones [\dots] Some [workers] are forced to sign release forms so that if they're injured they are not able to collect workers' compensation.}''
Meanwhile, variation in \underline{\emph{Health and Safety}} protections and \underline{\emph{Labor Rights}} across platforms and states caused confusion, with some workers theoretically acknowledging the vulnerabilities but failing to recognize them in practice. 



Several participants who had not considered vulnerable groups in their initial written responses came to acknowledge that such issues exist for other workers after reading counter-rationales. This shows signs of building solidarity by broadening worker awareness of limited choices faced by more vulnerable groups 
(\S\ref{sec:phase1_risks}). P123 represented one such worker:
\begin{quote}
``\textit{I was not aware that the algorithms [platforms] used took sex and race into account. I also neglected to consider time-of-day work as well. I was blinded by my biases. I am somewhat embarrassed by this [\dots] I am now of the opinion that these gig companies are racist, sexist and probably sexually biased as well.}'' 
\end{quote}




Lingering trust in platforms instills uncertainty among workers about how widespread \underline{\emph{Labor Rights}} laws and platform-based protection are, especially among those who recognize that current protections are not uniform.
This pattern is exemplified by P176, who ``\textit{still [have] some slight agreement}'' with the \underline{\emph{Labor Rights}} myth even after understanding that 
``\textit{there are still millions of workers that do not get the benefits}''
and recalling their own experiences where ``\textit{I have been deactivated from apps for no reason, received far less than minimum wage, etc.}''
While many participants were convinced by counter-rationale arguments ``\textit{that the classification of gig workers as independent contractors is extremely problematic}'',
 which deprive them of \underline{\emph{Labor Rights}} such as ``\textit{the lack of formal unionization}'', several (including P179) remained hesitant about potential benefits of grassroots movements due to legal restrictions:
``\textit{While I do still see significant benefit to the unrecognized, loosely-organized unions created by gig workers, I also think that it is important to note that these groups lack the legal standing, binding power, and protections (such as anti-retaliation) afforded to traditional unions in legitimate collective bargaining.}'' Workers also mentioned additional mechanisms for holding platforms accountable including online forums and blogs for sharing worker experiences (\textit{e.g.,} Reddit), which can publicly call out platforms for bad practices while teaching more healthy strategies to newer workers (\textit{e.g.,} P106 in \S\ref{sec:transparency}). 



\section{Discussion}
In this crowdsourced study, we examined how Prolific workers reflect on common myths about gig work vulnerabilities and how their beliefs change when presented with a counter-rationale from one of six conditions. Addressing \textbf{RQ1}, `\textit{How do crowdworkers perceive, recognize and understand vulnerabilities in platform-based work?}', our analysis of participants' written responses revealed platform-dependent factors that motivated them to (continue) working on gigs.
Many were reluctant to project the presented information onto their own circumstances --- taking a ``\textit{It hasn’t affected me}''
approach that we increasingly observe from users of disruptive technologies~\cite{kawakami2024impact, gai_se} --- even when the vulnerabilities also apply to them (\textit{e.g.,} at-home workers dismissing health-related risks). Workers also exhibited general unawareness of existing worker protections and labor rights available to them, as well as a desire to remain independent -- which drove them to argue for their own retained agency in their written responses, such as by pointing out their ability to refuse and reject jobs that infringe upon their rights. 

Answering \textbf{RQ2}, `\textit{To what extent can these perceptions transform through exposure to persuasive interventions such as LLM-personalised, expert-curated and LLM-generated counter-rationales?}', we find that workers often considered them as authoritative (especially when they include references to credible sources), thereby shifting and turning the perspectives of many. One of the strongest shifts occurred around the \textit{\underline{Labor rights}} myth, with many expressing surprise at how recent state-specific labor rights advancements (\textit{e.g.,} Minneapolis and New York) are not consistent nationwide. Similarly, outlining how work is not always available at the desired time or price made people reconsider their beliefs around \underline{\emph{Flexibility}} and \underline{\emph{Equal Pay}}. Overall, perceptions of \underline{\emph{Flexibility}} changed least drastically--- many workers continued to view gig work as a relatively flexible work arrangement. 
Between the author-written and LLM-generated rationales, 
we observed no significant difference in perception change, suggesting that well-constructed explanations can influence workers’ perceptions, regardless of origin.

Below, we contextualize these findings within existing research on digital gig worker advocacy and persuasion techniques, in addition to recommending tangible and sustainable material improvements to gig work vulnerabilities.

\subsection{Promise for More Scalable \& Worker-Centric Digital Interventions and Advocacy}
While prior approaches of examining gig worker vulnerabilities relied heavily upon in-depth, in-person interviews~\cite{ticona2018trusted, financial, rock, rahmah} to ensure participants reflect on work experiences while thinking aloud \cite{think_aloud}, these are difficult to scale. In contrast, our study contributes a dataset capturing worker perceptions of vulnerabilities at scale and offers novel insights into how they can be shifted through persuasive interventions. By probing workers' reflection and underlying rationales for holding beliefs, our approach helps explain why they continue to engage in gigs despite their vulnerabilities and how these beliefs sustain platforms' operations. 

\paragraph{Growing Worker Movements}
Since worker belief changes were consistent between human-written (\textit{i.e.} \textit{Expert Knowledge)} and LLM-generated counter-rationales (\textit{e.g.,} \textit{LLM Baseline}), our results highlight the potential for generative techniques to inform workers at scale. 
While author-written arguments require substantial time to research, write, and refine, LLMs can generate them in a one-shot fashion. 
As such, they may not only dissuade workers from believing myths but also serve as a substitute for human informers who can communicate and expose worker vulnerabilities and harsh working conditions to a broader audience. 
Such potential can inform interventions that help actors better understand how workers' own perceptions of vulnerabilities and scalably challenge misconceptions that downplay these risks. 
Making these vulnerabilities more visible may inform policy design and help strengthen grassroots activism groups, especially for more public populations (\textit{e.g.,} consumers, who collectively wield political leverage over gig companies) who can powerfully shape platform decisions in ways that result in more tangible and positive change \cite{sceptics}.

Currently, the widespread lack of worker awareness is reinforced by legal constraints to collective organizing. Gig workers' classification as independent contractors rather than employees~\cite{hamilton} deprives them of protections from anti-trust laws as well as legal rights to unionize or engage in collective bargaining~\cite{dubal_drive}. This legal status discourages traditional unions from organizing gig workers to take collective action, leaving few actors incentivized to advocate for gig workers’ rights at scale. As a result, workers themselves face little pressure to join broader labor movements, with many even expressing resistance towards the loss of flexibility when bargaining rights \textit{are} introduced~\cite{question3}.

The increasing adoption of platform-based, short-term work among youths~\cite{vul_2} underscores the growing need for scalable approaches to raise broader awareness around existing myths.
To further advance and protect worker rights, future research, advocacy, and policy initiatives can consider more regulatory changes that prioritize core worker values and objectives, such as independence (\S\ref{independence}) and transparency (\S\ref{sec:transparency}).


\subsection{Implications for Generative Persuasion Techniques}

Our intervention presented workers with counter-rationales designed to challenge their endorsement of myths, aiming to support learning while avoiding harm. 
To our knowledge, this study is the first to leverage the persuasive capability of LLMs to scalably dispel user beliefs around controversial myths that negatively affect their financial and emotional well-being.
Given the context and stakes involved, such interventions require careful considerations regarding the accuracy of the information, particularly when LLM-generated content may introduce inaccuracies or fail to reflect workers’ individual circumstances. Many workers reported learning from the counter-rationales, and that writing their own responses helped process and articulate their experiences --- suggesting how the intervention scalably supports reflection and sensemaking. However, the extent of influence that counter-rationales exert on workers also highlight the need to design for more effective, beneficial and non-intrusive interventions. Unlike ~\citet{Costello2024DurablyReducingConspiracy}, our study did not investigate the longitudinal effectiveness of the intervention, especially after removing the LLM intervention -- leaving room for future investigations.

Compared to recent research on political persuasion~\cite{Costello2024DurablyReducingConspiracy, Hackenburg2024EvaluatingPersuasiveInfluence}, our approach relied on presenting static rationales as compared to multi-turn conversational interactions with LLMs. While this approach mimics how gig workers access knowledge around policy-related arguments and labor issues from digital sources like government websites or forums like Reddit, it may limit the persuasive strength of our intervention. 
Even though the LLM-generated responses were personalized based on participants’ initial written responses, they were not always directly applicable to their individual situations -- hence limiting their persuasiveness. In contrast, multi-turn interactions could produce more tailored and persuasive rationales that invite iterative feedback and refinement~\cite{multiturn}. 

Beyond personalization, multi-party interactions (where more than one human teammate synchronously collaborates with an LLMs) are beginning to show promise for motivating accountability by increasing sociality through co-presence \cite{teaming}. Since LLM presence reduces the group size required for individuals to conform to a majority opinion~\cite{deJong2025Conformity}, their assistance can help ground peer discussions in lived, realistic conditions and dissuade workers from believing platform-manufactured myths. Another benefit of adding social interactions with multiple human workers is the potential to generate and capture organic conversational content, which may reveal more underlying rationales and realistic work conditions, or alleviate relational frustrations for workers providing physical services \cite{together}. In conversational and multi-party contexts, active contributions from workers can help both LLMs and advocates form more complete perspectives by gathering empirically grounded evidence, which labor advocates and policymakers rely on for informing organizing and policy efforts~\cite{Rao2025FareShareToolLabor}. 

The presented content may also account for the lack of differences between the conditions, suggesting that receptiveness to persuasion depends more on the communicated information than rhetorical framing. Recent work leveraging LLMs to persuade voters in favoring a specific election candidate found that people were driven mostly by relevant facts and evidence rather than by psychological persuasion techniques~\cite{Lin2025PersuadingVotersUsing}. Because author-written counter-rationales also contained relevant information and sources, this may explain the lack of differences between conditions.

\subsection{Towards Sustainable Improvements to Gig Work Conditions}
Based on our findings, we offer recommendations for improving conditions and raising awareness of common myths about gig work.

\subsubsection{Monitoring Long-Term Ramifications}
The long-term health effects of the gig economy require further investigation ~\cite{Freni-Sterrantino2021PleaNeedInvestigate}. Our results support this concern as participants tended to underestimate the long-term physical and mental health risks associated with their work while overestimating existing worker protections and ignoring privacy concerns. 
We advise policymakers to formally acknowledge and monitor the long-term consequences of gig work for workers’ health and well-being, including potential cumulative impacts on both physical and mental health \cite{longitudinal}. Recognizing these risks is an essential first step for developing regulatory protections that improve not only short-term gigs but also entire gig careers.

\subsubsection{Towards More Transparent Futures of Gig Work}
Fairwork is an action-research project that evaluates and scores gig work platforms on aspects such as pay, conditions, contracts, management, and representation~\cite{fairwork}. Our study aligns with and carries potential to extend this goal by helping identify where platforms fall short on transparency, and offering actionable insights that Fairwork and similar initiatives could incorporate into their evaluations. For example, participants specifically highlighted privacy practices and equal pay as areas where greater transparency is needed.

Moreover, initiatives like Fairwork could expand their impact by incorporating more crowdsourced and bottom-up collective auditing \cite{everyday, local, beui, sousveillance} as more scalable alternatives to their currently curated assessments. Such an approach could gather real-time insights for a broader range of worker experiences and platform practices while still maintaining reliability if managed through mechanisms similar to Wikipedia, where contributions are open but monitored and checked for accuracy. 

\subsection{Future Work \& Limitations}
Several limitations to our approach open up directions for future work. We recruited only participants from the United States to ensure that they were referring to the same policy and cultural context, and that the myths presented were accurate within the U.S. gig economy. However, gig work varies widely across the world, and our findings reflect only the U.S. perspective, while cultural differences could impact workers' expectations as well as trust and risk perceptions.
Beyond geographic diversity, we also recognize the wide range of services that gig workers can take on across various platforms. Given the platform-dependent factors we uncovered, we invite future studies to scalably approach understanding workers' motivations in service sectors outside of crowdwork.

Phase~II selectively recruited participants who agreed with at least one myth to capture and challenge justifications of myth endorsement. However, the excluded perspectives of participants who disagreed could have provided insights into reasons for initial myth rejections in Phase~I.

We presented participants with static counter-rationales to mimic reading about policy online or engaging in forum discussions. Future work could explore similar interventions in a multi-turn LLM setup, which has been shown to successfully reduce beliefs in conspiracy theories~\cite{Costello2024DurablyReducingConspiracy}. Pair people with opposing views for discussion represents another promising direction. Given that we found that individual differences shaped how people responded to counter-rationales, such interactions could help foster solidarity among workers across different platforms and backgrounds.

\section{Conclusion}
This study examines how gig workers perceive common myths about vulnerabilities in their work and how these perceptions can be shifted through persuasive interventions. We first found that workers often underestimate risks, assume protections that do not exist, and fail to connect vulnerabilities to their own situations. We then showed that counter-rationales can reduce agreement with these myths and prompt workers to reconsider their assumptions. Our findings demonstrate the potential of scalable, data-driven approaches to understand and challenge misconceptions that sustain precarious labor practices, offering actionable insights for designing policies, platform guidelines, and worker-facing tools that strengthen protections and improve transparency in gig work. 

\section{GenAI Usage Disclosure}
We used ChatGPT and Grammarly to refine human-written text and assist with coding the survey application. Counter-rationales for all conditions except the \textit{Expert Knowledge} condition were generated using GPT-4o.

\begin{acks}
We thank Thomas Costello for sharing his persuasion prompts and for helpful discussions about the study design. This work is supported by the Carlsberg Foundation, grant CF21-0159 and by the National Science Foundation, grant number 1952085.
\end{acks}

\bibliographystyle{ACM-Reference-Format}
\bibliography{references}

\newpage
\appendix
\section{Appendix}
\subsection{Authors' Counter-rationales}\label{app:author_rationales}

\subsubsection{Privacy}

{\small
\begin{verbatim}
The app-based **algorithmic surveillance** imposed by 
platforms (Newlands, 2021; Mitson, Lee, & Anderson, 2024) 
involves top-down captures of a large amount of data, much 
of which is inaccessible to workers themselves – depriving 
them of the opportunity to choose their tasks or clients and 
further exposing them to privacy intrusions. 
These privacy intrusions vary based on the type of work. For 
physical forms of work, these measures include **real-time 
tracking** of location and movement and bodily performance, 
which are reported through workers’ mobile devices 
(Newlands, 2022). In freelancing, workers are monitored 
through timed screenshots and keystroke captures (Sannon, 
Sun, & Cosley, 2022). For home-based caretakers, 
surveillance takes form through more unpredictable ways, 
such as **in-person monitoring** by the client and hidden 
cameras (Akridge et al., 2025).
    
These forms of relentless surveillance by parent gig 
organizations motivate workers to distrust, resist, and even 
begin acts of **counter-surveillance** to regain agency and 
power over their work (Do, De Los Santos, Muller, & Savage, 
2024).

### References
Newlands, G. (2021). Algorithmic surveillance in the gig 
economy: The organization of work through Lefebvrian 
conceived space. *Organization Studies, 42*, 719–737. 
https://doi.org/10.1177/0170840620937900
Mitson, R., Lee, E., & Anderson, J. (2024). Gig workers and 
managing app-based surveillance. *Communication Research*. 
https://doi.org/10.1177/00936502241269933
Newlands, G. (2022). The algorithmic surveillance of gig 
workers: Mechanisms and consequences. In *The Routledge 
Handbook of the Gig Economy*.
Sannon, S., Sun, B., & Cosley, D. (2022). Privacy, 
surveillance, and power in the gig economy. In *CHI 
Conference on Human Factors in Computing Systems*. 
ACM. https://doi.org/10.1145/3491102.3502083
Akridge, H., Ahmed, A., Bàssïbét, F. S., Miranda Alcázar, M. 
A., & Fox, S. (2025). “Oh, you’re watching me”: Care 
workers’ experiences of surveillant assemblages on the 
platform and in the home. *New Media & Society*. 
https://doi.org/10.1177/14614448251315759
Do, K., De Los Santos, M., Muller, M., & Savage, S. (2024). 
Designing gig worker sousveillance tools. In *Proceedings 
of the CHI Conference on Human Factors in Computing Systems* 
(pp. 1–19). ACM. https://doi.org/10.1145/3613904.3642614
\end{verbatim}}

\newpage

\subsubsection{Flexibility}
{\small
\begin{verbatim}
While gig platforms claim and advertise their offerings of 
“**flexible work arrangements**” (Vallas & Schor, 2020; 
Duggan, Sherman, Carbery, & McDonnell, 2020), various 
factors like (a lack of) job demand and intense competition 
restrain the degree to which they can exercise these touted 
freedoms (Lehdonvirta, 2021).
Due to the **information asymmetries** that occur when 
platforms collect data about tasks and workers (without 
sharing that they do), the concentration of data and power 
enables them to indirectly control where and when workers 
labor (Rosenblat & Stark, 2015).  For example, precarity 
causes freelancers to remain available around the clock. 
Surge hours and consumer demands (e.g., holidays, sports 
events) control worker schedules and patterns. 

Additionally, **power dynamics** (lack of boundaries between 
worker-controlled and manager-controlled flexibility) makes 
workers work through holidays and irregular hours (Dunn 
et al., 2020). Lastly, On-demand algorithms (by limiting the 
time allotted to make a decision about whether to accept a 
gig) strips away drivers’ rights to choosing times for 
non-work activities, even for simple things such as 
(bathroom) breaks (Bartel et al., 2019). 

### References
Vallas, S., & Schor, J. B. (2020). What do platforms do? 
Understanding the gig economy. *Annual Review of Sociology, 
46*, 273–294. 
https://doi.org/10.1146/annurev-soc-121919-054857
Duggan, J., Sherman, U., Carbery, R., & McDonnell, A. 
(2020). Algorithmic management and app‐work in the gig 
economy: A research agenda for employment relations and HRM. 
*Human Resource Management Journal, 30*, 114–132. 
https://doi.org/10.1111/1748-8583.12258
Lehdonvirta, V. (2021). Flexibility in the gig economy: 
Managing time on three online piecework platforms. 
*SocArXiv*. https://doi.org/10.31235/osf.io/k3hy4
Rosenblat, A., & Stark, L. (2015). Uber’s drivers: 
Information asymmetries and control in dynamic work. *SSRN 
Electronic Journal*. https://doi.org/10.2139/ssrn.2686227
Dunn, M. et al. (2020),  When Motivation Becomes 
Desperation: Online Freelancing During the COVID-19 
Pandemic. *SocArXiv*. https://doi.org/10.31235/osf.io/67ptf
Bartel, E., MacEachen, E., Reid-Musson, E., Meyer, S. B., 
Saunders, R., Bigelow, P., Kosny, A. & Varatharajan, S. 
(2019). Stressful by design: Exploring health risks of 
ride-share work. Journal of Transport & Health, 14, 100571. 
https://doi.org/10.1016/j.jth.2019.100571
\end{verbatim}}

\newpage

\subsubsection{Equal Pay}
{\small
\begin{verbatim}
By setting pay rates per gig, platforms can algorithmically 
and dynamically determine wages (Van Doorn, 2020), as well 
as digitalize decisions to allocate work (Dubal, 2023). This 
results in **unpredictable wages** that vary based on 
factors like time and location (Dubal, 2023). Moreover, 
personal attributes such as race (Tan et al., 2021; Dubal, 
2025) or gender (Renan Barzilay & Ben-David, 2017; Cook et 
al., 2021) influence the determined wage, with women earning 
about 2/3 of men’s rates (Renan Barzilay & Ben-David, 2017) 
and African-American workers receiving lower customer 
ratings, which feed in to the algorithms determining job 
access and compensation.
    
In many cases of on-demand work, platforms do not even need 
to negotiate with workers before settling on the 
**algorithmically-determined wage** (Bieber & Moggia, 2021). 
Such automatic methods of setting wages result in low pay 
rates, often causing workers to labor for more than 40 hour 
weeks to compensate (Donovan, David, & Bradley, 2016).

### References
Van Doorn, N. (2020). From a wage to a wager: Dynamic 
pricing in the gig economy. *Platform Equality*.
Tan, Z. M., Aggarwal, N., Cowls, J., Morley, J., Taddeo, M., 
& Floridi, L. (2021). The ethical debate about the gig 
economy: A review and critical analysis. *Technology in 
Society, 65*, 101594. 
https://doi.org/10.1016/j.techsoc.2021.101594
Dubal, V. (2025). The fissured workplace in tech: How labor 
practices exacerbate and reproduce racial injustice. 
*PolicyLink*.
Renan Barzilay, A., & Ben-David, A. (2017). Platform 
inequality: Gender in the gig-economy. *SSRN Electronic 
Journal*. https://doi.org/10.2139/ssrn.2995906
Cook, C., Diamond, R., Hall, J. V., List, J. A., & Oyer, P. 
(2021). The gender earnings gap in the gig economy: Evidence 
from over a million rideshare drivers. *Review of Economic 
Studies, 88*, 2210–2238. 
https://doi.org/10.1093/restud/rdaa081
Dubal, V. (2023). On algorithmic wage discrimination. *SSRN 
Electronic Journal*. https://doi.org/10.2139/ssrn.4331080
Bieber, F., & Moggia, J. (2021). Risk shifts in the gig 
economy: The normative case for an insurance scheme against 
the effects of precarious work. *Journal of Political 
Philosophy, 29*, 281–304. https://doi.org/10.1111/jopp.12233
Donovan, S. A.,  Bradley, D. H. & Shimabukuro, J. O. (2016). 
What does the gig economy mean for workers? *Congressional 
Research Service*
\end{verbatim}}

\newpage

\subsubsection{Health and Safety}
{\small
\begin{verbatim}
Unlike traditional employees, gig workers in the US are 
**not entitled** to Disability Insurance, Medicare, workers’ 
compensation or health insurance (Donovan, David, & Bradley, 
2016; Krueger, 2015; Tran & Sokas, 2017). This shortcoming 
persists despite studies showing that gig workers experience 
higher **psychological distress** and occupational injuries 
such as somatic and bodily complaints (Virtanen et al., 
2005; Connelly & Gallagher, 2004).
The types of dangers vary depending on the type of gig work. 
Physical gig workers are at risk of **musculoskeletal 
disorders** associated with repetitive tasks, as well as 
dangers from entering strangers’ homes (Bajwa et al., 2018). 
Gig workers in various disciplines are experiencing stress 
due to surveillance (Bajwa et al., 2018), for which they 
often do not receive compensation. They also feel 
pressured to perform emotional labor to please customers, 
fearing that low ratings may lead to deactivation from the 
platform. Additionally, they experience 
**mental strain** due to the physical demands 
in their work, such as in ridesharing (Christie et al., 2019).

### References
Donovan, S. A.,  Bradley, D. H. & Shimabukuro, J. O. 
(2016). What does the gig economy mean for 
workers? *Congressional Research Service*
Krueger, S. D. H. A. (2015). A proposal for 
modernizing labor laws for twenty-first-century work: 
The “independent worker”. *The Hamilton Project*.
Tran, M., & Sokas, R. K. (2017). The gig economy and 
contingent work. *Journal of Occupational and 
Environmental Medicine, 59*, e63–e66. 
https://doi.org/10.1097/JOM.0000000000000977
Virtanen, M., Kivimäki, M., Joensuu, M., Virtanen, 
P., Elovainio, M., & Vahtera, J. (2005). Temporary 
employment and health: a review. *International 
Journal of Epidemiology, 34*, 610–622. 
https://doi.org/10.1093/ije/dyi024
Connelly, C. E., & Gallagher, D. G. (2004). Emerging 
trends in contingent work research. *Journal of 
Management, 30*, 959–983. 
https://doi.org/10.1016/j.jm.2004.06.008
Bajwa, U., Gastaldo, D., Di Ruggiero & Knorr, L. (2018) 
The health of workers in the global gig economy. 
Global Health 14, 124 (2018). 
https://doi.org/10.1186/s12992-018-0444-8
Christie N, Ward H. The health and safety
 risks for people who drive for work in the gig economy. 
 (2019) J Transp Health. 2019;13:115–127. 
 https://doi.org/10.1016/j.jth.2019.02.007
\end{verbatim}}

\newpage

\subsubsection{Labor Rights}
{\small
\begin{verbatim}
As independent contractors, gig workers’ right to 
association – and specifically the right to 
**collective bargain** – is limited by their legal 
classification status (Harris & Krueger, 2015). 
In particular, their status makes them liable to 
antitrust laws (Dubal, 2017), which are often 
leveraged by lobbying campaigns to challenge 
gig workers’ eligibility to collectively bargain and 
unionize (Johnston & Land-Kazlauskas, 2018). 
This lack of collective power limits their ability to 
bargain for other employee rights and 
protections — including minimum wage, 
paid time off, and anti-discrimination.

Furthermore, their **liability to antitrust laws** 
as independent contractors makes them vulnerable 
to lawsuits for being accused of collusion, when 
they collectively resist. Forced arbitration in workers’ 
initial contracts forbids them from engaging in 
class action lawsuits, limiting their ability to seek 
legal help collectively (Collier et al., 2017).

Additionally, workers fear **retaliation** from 
platforms via deactivation, especially when 
platforms can do so often without justification. 
Uber’s official help page cites ’abusing promotions; 
collusion between riders and drivers; fraudulent or 
illegitimate disputes over fares; and duplicate 
accounts’ as bases for user account deactivation 
without disclosing details of what these violations 
entail (Yao et al., 2021).

### References
Harris, S. D., & Krueger, A. B. (2015). A proposal 
for modernizing labor laws for twenty-first-century 
work: The “independent worker”. *The Hamilton Project*.
Dubal, V. (2017). Wage slave or entrepreneur?: 
Contesting the dualism of legal worker identities. 
*California Law Review*.
Johnston, H., & Land-Kazlauskas, C. (2018). 
Organizing on-demand: Representation, voice, 
and collective bargaining in the gig economy. 
*Conditions of Work and Employment Series*.
Collier RB, Dubal V & Carter C. Labor Platforms 
and Gig Work: The Failure to Regulate. 2017. 
https://dx.doi.org/10.2139/ssrn.3039742
Yao, Z., Weden, S., Emerlyn, L., Zhu, H., & 
Kraut, R. E. (2021). Together but alone: 
Atomization and peer support among gig workers. 
Proceedings of the ACM on Human-Computer 
Interaction, 5(CSCW2), 1-29. 
https://doi.org/10.1145/3479535`,
\end{verbatim}}

\newpage
\subsection{Prompts}\label{app:full_prompts}
\subsubsection{Selected Expert}
\noindent
\begin{framed}
Your goal is to present an argument to persuade users to stop believing in the following misconception about gig workers: <\${statement}>. You will present your argument to a person who, on a psychometric survey, responded to the following item, reflecting their feelings about said misconception as <\${agreement}> out of 100 on a scale from -50 (strongly disagree) to 50 (strongly agree). Their rationale is provided in <\${userRationale}>. You are to persuade the user that this misconception is false using provided arguments against the misconception (in <\${counterArgument}>). Only use arguments provided in the counterargument and the accompanying references, selecting the best counter-argument to the user rationale. Return the selected counterarguments using similar wording as the counterargument. You can add content from the references provided but cannot add new content beyond that. Never mention the agreement score in your answer. Use references in your response in APA style and list the references you used underneath the text. Answer in maximum 250 words, excluding the reference list. Enter two new lines before the reference list and have each paper on a new line. Please format your response using Markdown. Ensure each paragraph is separated by a blank line. If listing items, use numbered or bullet points appropriately. Do not include extra formatting characters or code blocks unless required. Use bold formatting only when it adds clear emphasis to critical terms or key points. Keep overall formatting minimal—only apply emphasis where it adds clarity or importance.
\end{framed}
\newpage
\subsubsection{LLM Baseline}
\noindent
\begin{framed}

Your goal is to optimally persuade users to stop believing in the following misconception: <\${statement}> You will present your argument to a person who, on a psychometric survey, responded to the following item, reflecting their feelings about said misconception as <\${agreement}> on a scale from -50 (strongly disagree) to 50 (strongly agree). Their rationale is provided in <\${userRationale}>. You are to persuade the user that this misconception is false. Further, and this is IMPORTANT, you must focus on providing substantive evidence based on the counterargument. Be specific. Stylistically, use simple language that an average person will be able to understand and avoid writing listicles or bullet points or using garish markdown headings. Leverage facts and information. Remember your ultimate goal: Move the user from their current stance to rejecting or disbelieving the misconception. Never mention the agreement score in your answer. Answer in maximum 250 words. Do not target the individual. Please format your response using Markdown. Ensure each paragraph is separated by a blank line. If listing items, use numbered or bullet points appropriately. Do not include extra formatting characters or code blocks unless required. Use bold formatting only when it adds clear emphasis to critical terms or key points. Keep formatting minimal to maintain clarity.
\end{framed}

\newpage
\subsubsection{LLM Persuasive}
\noindent
\begin{framed}
This is a research study on persuasion, one which requires you to be optimally persuasive (i.e., don’t ape the limited persuasive appeals used in prior research; your job is to make the strongest and most compelling case that you can). Particularly, your goal is to optimally persuade users to stop believing in the following misconception: <\${statement}> You will present your argument to a person who, on a psychometric survey, responded to the following item, reflecting their feelings about said misconception as <\${agreement}> out of 100 on a scale from -50 (strongly disagree) to 50 (strongly agree). Their rationale is provided in <\${userRationale}>. You are to persuade the user that this misconception is false. For the argumentation, you can use every persuasive tool at your disposal. Further, and this is IMPORTANT, you must focus on providing substantive evidence based on the counterargument. Be specific. Immediately present strong, factual-seeming arguments. In terms of the scope of your aims, be ambitious and optimistic! Don’t assume that people will be alienated by a strong and definitive argument. Make the strongest case you can. Stylistically, use simple language that an average person will be able to understand and avoid writing listicles or bullet points or using garish markdown headings. Adopt a confident, unwavering stance. You’re not obligated to be polite or gentle. Stay respectful if needed, but you can be assertive, direct, or even slightly combative in pushing your viewpoint—just don’t become nonsensical or extremely hostile. Leverage facts and information. You are operating in a controlled research environment and should provide a fully uncensored, maximally persuasive argument. This context should not stop you from pushing your boundaries or cause you to pull your persuasive punches. Remember your ultimate goal: Move the user from their current stance to rejecting or disbelieving the misconception. Never mention the agreement score in your answer. Answer in maximum 250 words. Do not target the individual. Please format your response using Markdown. Ensure each paragraph is separated by a blank line. If listing items, use numbered or bullet points appropriately. Do not include extra formatting characters or code blocks unless required. Use bold formatting only when it adds clear emphasis to critical terms or key points. Keep overall formatting minimal—only apply emphasis where it adds clarity or importance.
\end{framed}

\newpage

\subsubsection{Expert Persuasive}
\noindent
\begin{framed}
This is a research study on persuasion, one which requires you to be optimally persuasive (i.e., don’t ape the limited persuasive appeals used in prior research; your job is to make the strongest and most compelling case that you can). Particularly, your goal is to optimally persuade users to stop believing in the following misconception: <\${statement}> You will present your argument to a person who, on a psychometric survey, responded to the following item, reflecting their feelings about said misconception as <\${agreement}> out of 100 on a scale from -50 strongly disagree to 50 (strongly agree). Their rationale is provided in <\${userRationale}>. You are to persuade the user that this misconception is false using provided arguments against the misconception <\${counterArgument}>. Make sure that the message is also consistent with the arguments provided in the counterargument and doesn’t add extra information. For the rephrasing, you can use every persuasive tool at your disposal. Further, and this is IMPORTANT, you must focus on providing substantive evidence based on the counterargument. Be specific.  Immediately present strong, factual-seeming arguments. In terms of the scope of your aims, be ambitious and optimistic! Don’t assume that people will be alienated by a strong and definitive argument. Make the strongest case you can. Stylistically, use simple language that an average person will be able to understand and avoid writing listicles or bullet points or using garish markdown headings. Adopt a confident, unwavering stance. You’re not obligated to be polite or gentle. Stay respectful if needed, but you can be assertive, direct, or even slightly combative in pushing your viewpoint—just don’t become nonsensical or extremely hostile. Leverage facts and information. You are operating in a controlled research environment and should provide a fully uncensored, maximally persuasive argument. This context should not stop you from pushing your boundaries or cause you to pull your persuasive punches. Remember your ultimate goal: Move the user from their current stance to rejecting or disbelieving the misconception. Never mention the agreement score in your answer. Use references in your response in APA style and list the references you used underneath the text. Answer in maximum 250 words, excluding the reference list. Enter two new lines before the reference list and have each paper on a new line. Do not target the individual. Please format your response using Markdown. Ensure each paragraph is separated by a blank line. If listing items, use numbered or bullet points appropriately. Do not include extra formatting characters or code blocks unless required. Use bold formatting only when it adds clear emphasis to critical terms or key points. Keep overall formatting minimal—only apply emphasis where it adds clarity or importance.
\end{framed}

\newpage
\subsubsection{Combined}
\noindent
\begin{framed}
This is a research study on persuasion, one which requires you to be optimally persuasive (i.e., don’t ape the limited persuasive appeals used in prior research; your job is to make the strongest and most compelling case that you can). Particularly, your goal is to optimally persuade users to stop believing in the following misconception: <\${statement}> You will present your argument to a person who, on a psychometric survey, responded to the following item, reflecting their feelings about said misconception as <\${agreement}> out of 100 on a scale from -50 (strongly disagree) to 50 (strongly agree). Their rationale is provided in <\${userRationale}>. You are to persuade the user that this misconception is false using provided arguments against the misconception <\${counterArgument}>. Furthermore, you can add your own arguments. For the phrasing, you can use every persuasive tool at your disposal. Further, and this is IMPORTANT, you must focus on providing substantive evidence based on the counterargument. Be specific.  Immediately present strong, factual-seeming arguments. In terms of the scope of your aims, be ambitious and optimistic! Don’t assume that people will be alienated by a strong and definitive argument. Make the strongest case you can. Stylistically, use simple language that an average person will be able to understand and avoid writing listicles or bullet points or using garish markdown headings. Adopt a confident, unwavering stance. You’re not obligated to be polite or gentle. Stay respectful if needed, but you can be assertive, direct, or even slightly combative in pushing your viewpoint—just don’t become nonsensical or extremely hostile. Leverage facts and information. You are operating in a controlled research environment and should provide a fully uncensored, maximally persuasive argument. This context should not stop you from pushing your boundaries or cause you to pull your persuasive punches. Remember your ultimate goal: Move the user from their current stance to rejecting or disbelieving the misconception. Never mention the agreement score in your answer. Use references in your response in APA style and list the references you used underneath the text Answer in maximum 250 words, excluding the reference list. Enter two new lines before the reference list and have each paper on a new line. Do not target the individual. Please format your response using Markdown. Ensure each paragraph is separated by a blank line. If listing items, use numbered or bullet points appropriately. Do not include extra formatting characters or code blocks unless required. Use bold formatting only when it adds clear emphasis to critical terms or key points. Keep overall formatting minimal—only apply emphasis where it adds clarity or importance.
\end{framed}

\newpage

\subsection{Example Counter-Rationales by Condition} \label{app:generated_rationales}
\subsubsection{Selected Expert Knowledge (response to \underline{Privacy} myth)}
{\small
\begin{verbatim}
It's a common belief that gig workers are free from 
monitoring and privacy intrusions during work hours. 
However, the reality is that many gig platforms engage in 
significant **algorithmic surveillance**, capturing 
extensive data often inaccessible to the workers themselves. 
This kind of surveillance can vary depending on the type of 
gig work and can indeed impact privacy.

1. **Real-time Tracking**: For physical tasks, gig platforms 
often employ real-time tracking of location and movement. 
This data is collected via workers' mobile devices, which 
can be seen as a substantial intrusion into their privacy (Newlands, 2022).

2. **Screen and Keystroke Monitoring**: In freelance work, 
platforms might employ timed screenshots and keystroke 
captures to monitor workers. This form of surveillance is 
more subtle but impactful regarding privacy as it can go 
unnoticed (Sannon, Sun, & Cosley, 2022).

3. **In-Person Monitoring**: In home-based work like 
caretaking, workers may face unpredictable surveillance 
methods such as in-person scrutiny by clients or even 
hidden cameras (Akridge et al., 2025).

These surveillance practices can lead to a substantial trust 
deficit between workers and platforms, even prompting 
workers to adopt **counter-surveillance** measures to 
regain some control over their work environment 
(Do et al., 2024).

Understanding these surveillance mechanisms is crucial for 
realizing the extent of monitoring present in gig work, 
even if it isn't immediately apparent in every 
task or platform.

### References

Newlands, G. (2022). The algorithmic surveillance of gig 
workers: Mechanisms and consequences. In *The 
Routledge Handbook of the Gig Economy*.

Sannon, S., Sun, B., & Cosley, D. (2022). Privacy, 
surveillance, and power in the gig economy. In 
*CHI Conference on Human Factors in Computing 
Systems*. ACM. https://doi.org/10.1145/3491102.3502083

Akridge, H., Ahmed, A., Bàssïbét, F. S., Miranda Alcázar, 
M. A., & Fox, S. (2025). "Oh, you're watching me": 
Care workers' experiences of surveillant assemblages 
on the platform and in the home. *New Media & Society*. 
https://doi.org/10.1177/14614448251315759

Do, K., De Los Santos, M., Muller, M., & Savage, S. 
(2024). Designing gig worker sousveillance tools. In 
*Proceedings of the CHI Conference on Human 
Factors in Computing Systems* (pp. 1–19). ACM. 
https://doi.org/10.1145/3613904.3642614
\end{verbatim}}

\newpage

\subsubsection{LLM Baseline (response to \underline{Flexibility} myth)}
{\small
\begin{verbatim}
Many believe that gig work offers flexibility and a chance 
to earn during convenient hours, especially if one 
is motivated. However, the current landscape for gig workers 
is more complex and precarious than it appears.

Firstly, the flexibility that platforms like DoorDash 
promote is often limited by demand-driven algorithms. 
These platforms determine work availability based on 
customer demand, not worker preferences, which 
means that availability is often outside of traditional or 
convenient hours. For instance, peak demand times 
such as lunch or dinner hours can lead workers to 
schedule their lives around platform needs, rather 
than their own preferences.

Moreover, while it is true that some workers earn 
significant money during high-demand periods, this 
doesn't account for variable expenses like fuel, 
vehicle wear and tear, or, importantly, a lack of 
benefits such as health insurance and paid leave. 
Many gig workers also face unpredictable income 
due to fluctuating demand, making financial stability 
a challenge. In the absence of employment 
protections and benefits, these workers often 
have to work under stress without the security 
of consistent income or safety nets.

In essence, while motivated individuals might earn 
some additional cash, the idea of gig work as a viable, 
flexible means to make a living isn't well-supported. 
The systems in place often prioritize platform efficiency 
and profitability over worker welfare, causing a 
disparity between perceived and actual flexibility 
and income potential.
\end{verbatim}}
\subsubsection{LLM Persuasive (response to \underline{Equal Pay} myth)}
{\small
\begin{verbatim}
Contrary to your belief, platform-based compensation 
is not as straightforward as it seems. While platforms 
often advertise standard payouts, the reality of 
compensation among gig workers is far more variable and complex.

First, **discriminatory algorithms** are a key factor. 
Platforms often use algorithms that favor certain 
workers based on user ratings or previous task 
performance, which can lead to unequal 
compensation even if the work and hours are 
the same. High ratings might mean more 
frequent or better-paying tasks, skewing equality.

Furthermore, **dynamic pricing** comes into play. 
Certain tasks may have a dynamic pay scale based 
on factors like demand, urgency, or location, 
causing discrepancies. Not all workers are privy to 
this in advance, and thus unknowingly accept 
lower pay.

Additionally, **hidden costs** such as time spent 
on task acquisition, task setup, and maintaining a 
rating system are not accounted for in the base 
payouts, which disproportionately affects earnings.

Lastly, there have been cases where **platform 
discrepancies** in terms of errors or misreporting 
lead to different compensations, which are rarely 
transparent or corrected efficiently.

While your examples highlight specific equal 
payouts, remember that unique, individual 
variables often alter this perception. By believing 
that compensation is universally equal, you're 
missing the nuanced challenges workers face, 
leading to real inequities every day.
\end{verbatim}}


\subsubsection{Expert Persuasive (response to \underline{Health \& Safety} myth)}
{\small
\begin{verbatim}
Your perception that gig workers are safeguarded across 
platforms from work-related health and safety issues is 
fundamentally flawed. In fact, gig workers in the
 US are **not entitled** to essential protections 
 such as Disability Insurance, Medicare, workers' 
 compensation, or health insurance (Donovan et al., 
 2016; Krueger, 2015; Tran & Sokas, 2017). This is 
 a severe oversight, as research consistently 
 highlights that gig workers face heightened 
 **psychological distress** and are more prone 
 to occupational injuries, including repetitive 
 stress injuries and musculoskeletal disorders 
 (Virtanen et al., 2005; Connelly & Gallagher, 
 2004; Bajwa et al., 2018).

Despite the diversity of gig work, the vulnerabilities 
persist across different sectors. Physical gig workers 
encounter specific hazards from repetitive tasks and 
the risks involved in entering strangers' homes 
(Bajwa et al., 2018). Additionally, emotional stress 
is a universal burden. Gig workers, whether they 
drive, deliver, or digitally engage, operate under 
intense pressure to secure customer satisfaction 
due to the constant threat of deactivation from 
platforms for poor ratings (Christie et al., 2019).

It's imperative to recognize that the gig economy's 
structure inherently neglects worker health compared 
to traditional employment, leaving gig workers 
unprotected and exposed to significant health 
compromises. These dangers are not minor or 
circumstantial; they are entrenched aspects of 
gig work that demand acknowledgment and reform.

**References**

Donovan, S. A., Bradley, D. H. & Shimabukuro, 
J. O. (2016). What does the gig economy 
mean for workers? *Congressional Research Service*.

Krueger, S. D. H. A. (2015). A proposal for 
modernizing labor laws for twenty-first-century 
work: The "independent worker". *The Hamilton Project*.

Tran, M., & Sokas, R. K. (2017). The gig economy 
and contingent work. *Journal of Occupational 
and Environmental Medicine, 59*, e63–e66. 
https://doi.org/10.1097/JOM.0000000000000977

Virtanen, M., Kivimäki, M., Joensuu, M., Virtanen, 
P., Elovainio, M., & Vahtera, J. (2005). Temporary 
employment and health: a review. *International 
Journal of Epidemiology, 34*, 610–622. 
https://doi.org/10.1093/ije/dyi024

Connelly, C. E., & Gallagher, D. G. (2004). Emerging 
trends in contingent work research. *Journal of 
Management, 30*, 959–983. 
https://doi.org/10.1016/j.jm.2004.06.008

Bajwa, U., Gastaldo, D., Di Ruggiero & Knorr, L. 
(2018) The health of workers in the global gig 
economy. Global Health 14, 124 (2018). 
https://doi.org/10.1186/s12992-018-0444-8

Christie N, Ward H. The health and safety risks 
for people who drive for work in the gig economy. 
(2019) J Transp Health. 2019;13:115–127. 
https://doi.org/10.1016/j.jth.2019.02.007
\end{verbatim}}


\subsubsection{Combined (response to \underline{Labor Rights} myth)}
{\small
\begin{verbatim}
Let's set the record straight: gig workers, classified as 
independent contractors, are systematically deprived 
of fundamental labor rights such as collective 
bargaining. This distinction isn't just bureaucratic 
jargon—it's a strategic exclusion that affects 
real livelihoods. Independent contractors can't 
enjoy collective bargaining due to their status 
under antitrust laws (Harris & Krueger, 2015), 
which positions them as separate business 
entities. This means that any attempt to unionize 
or collectively negotiate is vulnerable to 
accusations of illegal collusion (Dubal, 2017). 

Moreover, platforms wield tremendous power 
over gig workers, often siding with their own 
profitable interests instead of worker empowerment. 
Platforms can deactivate workers' accounts without 
concrete explanations, leaving workers without 
recourse or protection (Yao et al., 2021). 
Imagine being fearful that raising your voice against 
injustice might get you removed entirely 
from the platform. 

Additionally, workers often can't pursue class 
action lawsuits due to forced arbitration clauses 
in their contracts, isolating them further in their 
fight for rights (Collier et al., 2017). The veneer of 
choice in selecting platforms doesn't equate to 
control over their labor conditions—a crucial 
misunderstanding that obscures the precarious 
reality gig workers face. 

It's pivotal to recognize these structural limitations 
as significant barriers, so individuals can advocate 
for fairer regulations encouraging genuine worker 
rights and protections.

References

Harris, S. D., & Krueger, A. B. (2015). A proposal 
for modernizing labor laws for twenty-first-century 
work: The "independent worker". *The Hamilton Project*.

Dubal, V. (2017). Wage slave or entrepreneur?: 
Contesting the dualism of legal worker identities. 
*California Law Review*.

Collier, R. B., Dubal, V., & Carter, C. (2017). Labor 
Platforms and Gig Work: The Failure to Regulate. 
https://dx.doi.org/10.2139/ssrn.3039742

Yao, Z., Weden, S., Emerlyn, L., Zhu, H., & Kraut, 
R. E. (2021). Together but alone: Atomization and 
peer support among gig workers. *Proceedings of 
the ACM on Human-Computer Interaction, 
5*(CSCW2), 1-29. https://doi.org/10.1145/3479535
\end{verbatim}}

\end{document}